\documentclass[12pt]{article}

\usepackage{newtxtext,newtxmath}

\usepackage{graphicx}

\usepackage[letterpaper,margin=1in]{geometry}

\renewenvironment{abstract}
	{\quotation}
	{\endquotation}

\makeatletter
\renewcommand{\fnum@figure}{\textbf{Figure \thefigure}}
\renewcommand{\fnum@table}{\textbf{Table \thetable}}
\makeatother

\usepackage{scicite}

\usepackage{url}

\def\scititle{A pulsar-helium star compact binary system formed by common envelope evolution
}
\title{\bfseries \boldmath \scititle}

\author{ 
Z. L. Yang$^{1,2}$,
J. L. Han$^{1,2\ast}$, 
D. J. Zhou$^{1,2}$,  
W. C. Jing$^{1,2}$,
W. C. Chen$^{3}$,  \and 
T. Wang$^{1,2}$, 
X. D. Li$^{4,5}$,
S. Wang$^{1,2}$, 
B. Wang$^{2,6}$, 
H. W. Ge$^{2,6}$, \and
Y. L. Guo$^{4,5}$, 
L. H. Li$^{2,6}$,  
Y. Shao$^{4,5}$, 
J. F. Liu$^{1,2}$,
W.~Q. Su$^{1,2}$, 
L. G. Hou$^{1,2}$, \and
W. J. Huang$^{7}$,
J. C. Jiang$^{1}$, 
P. Jiang$^{1}$, 
J. H. Sun$^{1}$,
B. J. Wang$^{1}$, 
C. Wang$^{1}$, \and
H. G. Wang$^{9,10}$, 
J. B. Wang$^{11}$,  
N. Wang$^{11}$,  
P. F. Wang$^{1}$, 
S. Q. Wang$^{11}$, \and
H. Xu$^{1}$,  
J. Xu$^{1}$, 
R. X. Xu$^{8}$,  
W. M. Yan$^{11}$,  
Y. Yan$^{1,2}$,
X. P. You$^{12}$,\and
D. J. Yu$^{1}$,
Z. S. Yuan$^{1}$, 
C. F. Zhang$^{1}$ 
\and 
\small$^{1}$ National Astronomical Observatories, Chinese Academy of Sciences, Beijing, China \and 
\small$^{2}$School of Astronomy and Space Sciences, University of Chinese Academy of Sciences, \small Beijing, China  \and 
\small$^{3}$ School of Science, Qingdao University of Technology, Qingdao, China  \and 
\small$^{4}$ School of Astronomy and Space Science,  Nanjing University, Nanjing, China \and 
\small$^{5}$ Key Laboratory of Modern Astronomy and Astrophysics, Nanjing University, Nanjing, China \and 
\small$^{6}$ Yunnan Observatories, Chinese Academy of Sciences, Kunming, China \and 
\small$^{7}$ School of Physics and Astronomy, Sun Yat-sen University, Zhuhai, China \and 
\small$^{8}$  Department of Astronomy, Guangzhou University, Guangzhou, China \and 
\small$^{9}$  National Astronomical Data Center, Great Bay Area, Guangzhou, China \and 
\small$^{10}$  Xinjiang Astronomical Observatory, Chinese Academy of Sciences, Urumqi, China \and 
\small$^{11}$ Department of Astronomy, Peking University, Beijing 100871, China \and 
\small$^{12}$ School of Physical Science and Technology, Southwest University, Chongqing, China \and 
\small$^\ast$Corresponding author. E-mail: hjl@nao.cas.cn
}

\date{}
\begin{document} 

\baselineskip24pt

\maketitle 
\begin{abstract} \bfseries \boldmath 
A stellar common envelope occurs in a binary system when the atmosphere of an evolving star expands to encompass an orbiting companion object. Such systems are predicted to evolve rapidly, ejecting the stellar envelope and leaving the companion in a tighter orbit around a stripped star. We used radio timing to identify a pulsar, PSR J1928+1815, with a spin period of 10.55 ms in a compact binary system with an orbital period of 3.60 hours. The companion star has 1.0 to 1.6 solar masses, eclipses the pulsar for about 17\% of the orbit, and is undetected at other wavelengths, so it is most likely a stripped helium star. We interpret this system as having recently undergone a common envelope phase, producing a compact binary.
\end{abstract} 

Millisecond pulsars - neutron stars with spin periods of less than about 30 milliseconds - acquired their rapid rotation by accreting material from a binary companion star 
\cite{Alpar+1982Natur.300..728A}.
It is unclear how binary systems that contain pulsars form, because there are multiple potential evolutionary pathways \cite{Han+2020RAA....20..161H,Tauris+2023pbse.book.....T}. Depending on the initial masses of the component stars, the evolution of a binary system can involve multiple mass-transfer phases, one or more common envelope evolution, and one or two supernova explosions. A common envelope phase occurs when one star expands sufficiently for its atmosphere to encompass the companion \cite{Paczynski+1976IAUS...73...75P}. This causes the orbit to shrink rapidly, within $10^3$ years, leading to either a merger, or the ejection of the envelope leaving a stripped star in a very compact binary orbit \cite{Paczynski+1976IAUS...73...75P,Iben+1993PASP..105.1373I,Ivanova+2013A&ARv..21...59I}. If the companion object is a neutron star, this process is expected to produce a binary system consisting of a stripped helium star and a recycled pulsar, but there is little observational evidence for such systems. Stripped helium stars have been observed in the Magellanic Clouds (nearby dwarf galaxies), some of them in binaries with unidentified companions \cite{Drout+2023Sci...382.1287D}.

In such a binary formed through the common envelope evolution, the helium star can eclipse the pulsar. Some previously known millisecond pulsars are eclipsed by their companions in compact orbits with an orbital period $P_{\rm orb} \lesssim 1$~day  \cite{Chen_2013, Pan+2021ApJ...915L..28P,  Roberts+2013IAUS..291..127R, Smith+2023ApJ...958..191S}, devouring the companions as spiders. They are further classified as redbacks and black widows according to companion masses. Black widows have a companion of brown dwarf star or white dwarf remnant with a mass of much less than 0.1 solar~masses ($M_\odot$), highly ablated by the pulsar wind. Redback pulsars have companions of irritated main sequence stars with masses of 0.1 to 0.4~$M_\odot$. None of the spider pulsar companions is a helium star.

\paragraph*{Radio observations of PSR~J1928+1815}\label{sec2}

PSR~J1928+1815 (coordinates in Table 1) was previously identified using the Five-hundred-meter Aperture Spherical radio Telescope (FAST) \cite{Nan+2006ScChG..49..129N} during the FAST Galactic Plane Pulsar Snapshot survey \cite{Han+2021RAA....21..107H}. The pulsar has a spin period $P$ of 10.55~ms, indicating recent mass transfer, and is in a binary system with an orbital period of 3.60 hours, which accelerates the pulsar by 73.0 $\rm m\,s^{-2}$ \cite{Han+2021RAA....21..107H}. We performed 4.5 years of follow-up pulsar timing observations using FAST with its $L$-band receiver working at radio frequency range of 1.0 to 1.5 GHz \cite{mm, Jiang+2020RAA....20...64J}. The times of arrival (TOAs) of pulses were fitted with a binary orbit model \cite{mm} to produce phase-coherent timing solutions (Table~\ref{timing solution}). hat the binary system has eccentricity $e<3\times10^{-5}$, indicating a circular
orbit. The spin period derivative is $3.63\times10^{-18}$ s~s$^{-1}$, which is high compared to with that of other millisecond pulsars, indicating rapid energy loss due to spin down of the pulsar. This implies that the PSR~J1928+1815 binary system has recently experienced mass transfer.

From the observationally measured parameters, we determined the mass function
\begin{equation}
    f(M_{\rm p},M_{\rm c},i)\equiv(M_{\rm c}\sin i)^3/(M_{\rm p}+M_{\rm c})^2=\frac{4\pi^2}{G}{x^3}/{P_{\rm orb}^2},
	\label{m_f}
\end{equation}
where $M_{\rm p}$ and $M_{\rm c}$ are the masses of the pulsar and companion, $i$ is the orbital inclination angle, $G$ is the gravitational constant. Variable $x$ is the length in light second of the projected semi-major axis of the orbit which can be measured from the pulse time delay. We found that the mass function of PSR J1928+1815 is 0.2342~$M_\odot$, which sets a strict lower limit on the mass of the companion.

\begin{figure}
\centering
     \includegraphics[width=0.55\columnwidth]{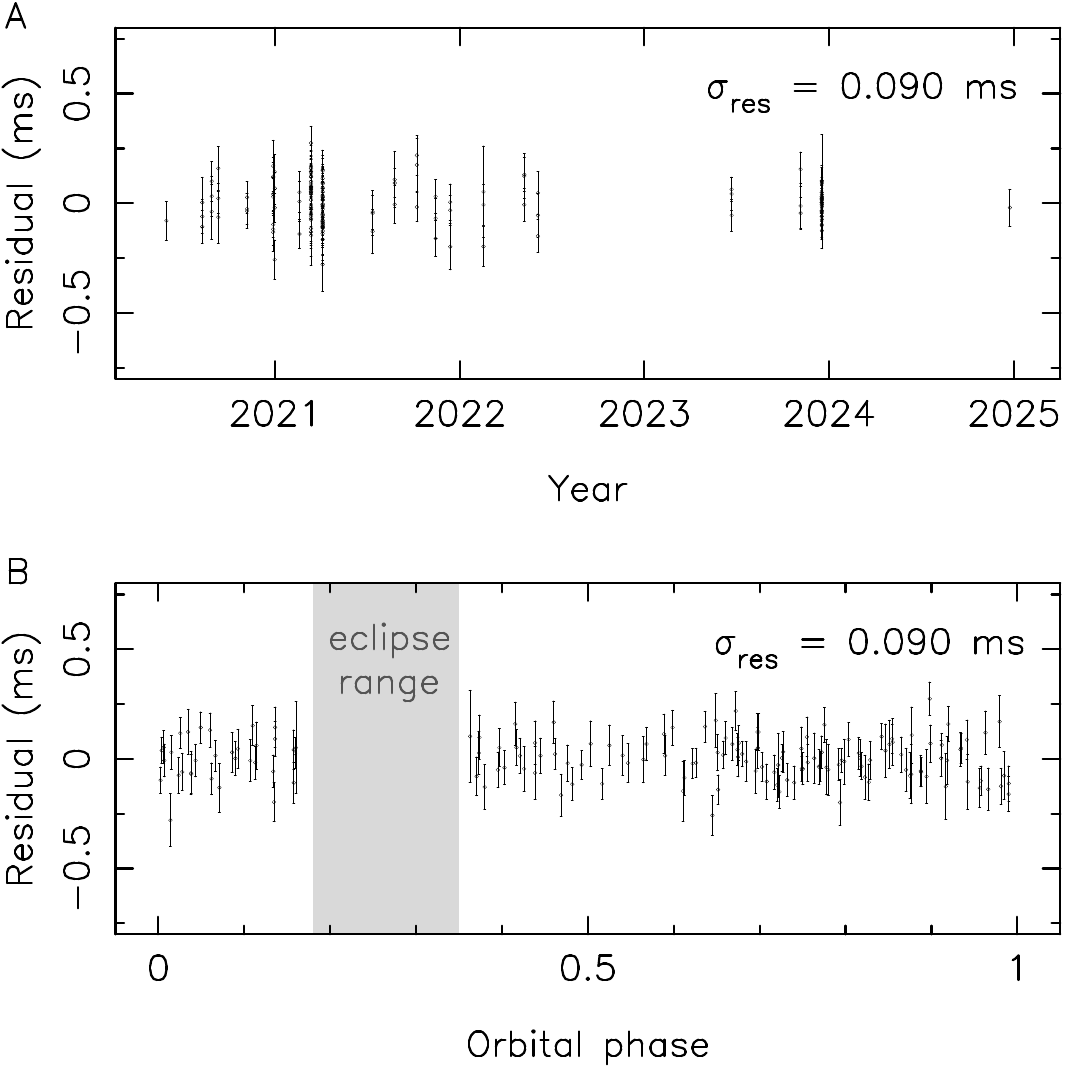}\\
     \includegraphics[width=0.55\columnwidth]{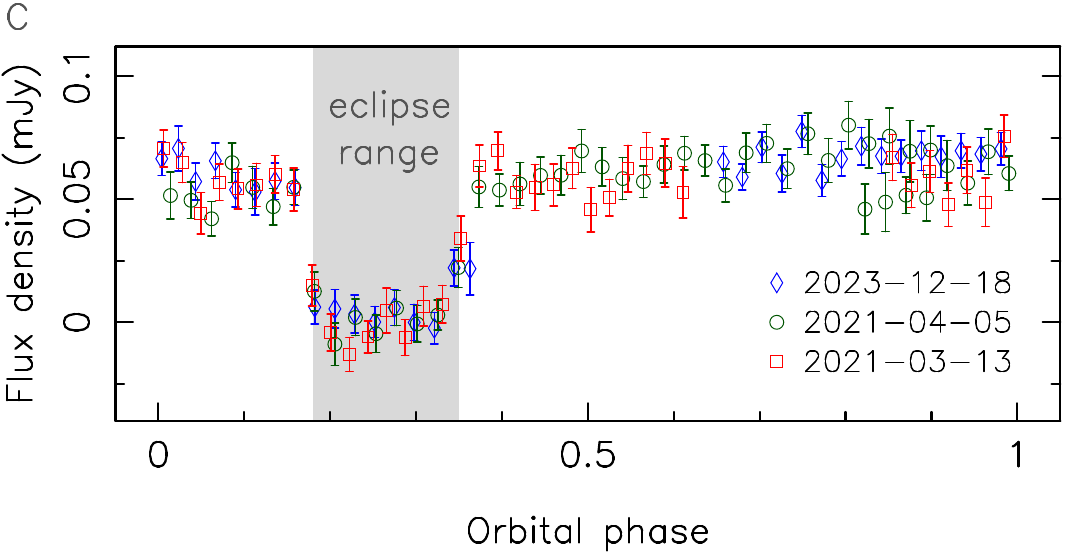}\\
    \caption{{\bf Pulsar timing and eclipses of PSR~J1928+1815.} (A) Residuals between the pulse TOAs measured by FAST and the timing model (Table~\ref{timing solution}), as a function of observation date. All TOAs were measured from 5-minute observation segments. The weighted root-mean-square of the residuals $(\sigma_{\rm res})$ is indicated. (B) Same as (A), but as a function of orbital phase. (C) Pulsed flux densities in millijansky (mJy; 1 mJy = $10^{-29}$~W\,m$^{-2}$\,Hz$^{-1}$) measured during three longer tracking FAST observations (colored symbols, see legend) as functions of orbital phase. In (B) and (C), shading indicates phases when no (or very weak) pulses were detected, which we interpret as being due to eclipses by the companion when it is in the front of the pulsar at the orbital phase around $\phi_0=0.25$. All error bars show $\pm 1\sigma$ uncertainties. }
    \label{Timing result & flux vs. orbital phase}
\end{figure}

We performed three longer observations using FAST to cover the full orbital period \cite{mm}. These show regular long gaps in radio emission during the orbital phase range of 0.18 to 0.35 (see Fig.~\ref{Timing result & flux vs. orbital phase}C and fig.~\ref{time-phase}). This is around the superior conjunction, at orbital phase 0.25, when our timing solution indicates that the pulsar is located directly on the far side of the companion from Earth. This indicates that the gaps are caused by eclipses of the pulsar by the companion.

\paragraph*{Constraints on the companion} 
The mass of PSR~J1928+1815$'$s binary companion is constrained by the mass function presented in Eq.~\ref{m_f}. All previously measured pulsar masses s (with precision $<$0.2~$M_\odot$) are in the range of 1.1 to 2.2~$M_\odot$ \cite{Tauris+2023pbse.book.....T}, if PSR~J1928+1815 has the lowest mass in that range, that is, 1.1~$M_\odot$, then Eq.~\ref{m_f} indicates the companion mass must be $>$1.0~$M_\odot$. This is much more massive than the companions of spider pulsars (fig.~\ref{Pb-m2}). Adopting a higher pulsar mass would imply an even more massive companion.

If the companion was a massive white dwarf or a neutron star,  it would not cause an eclipse as long as that which we observed for PSR~J1928+1815. We attributed the regular eclipses (Fig.~\ref{Timing result & flux vs. orbital phase}, B and C) to the outflowing material (stellar wind) surrounding the companion; to block the radio emission, the companion’s stellar wind must be much denser than the stellar
wind that a white dwarf or a neutron star would produce. The observed eclipses also indicate a highly inclined orbit (close to edge on), such that the companion’s stellar wind intersects the line of sight between Earth and the pulsar during part of the orbit.

\begin{figure*}
\centering
     \includegraphics[width=0.8\columnwidth]{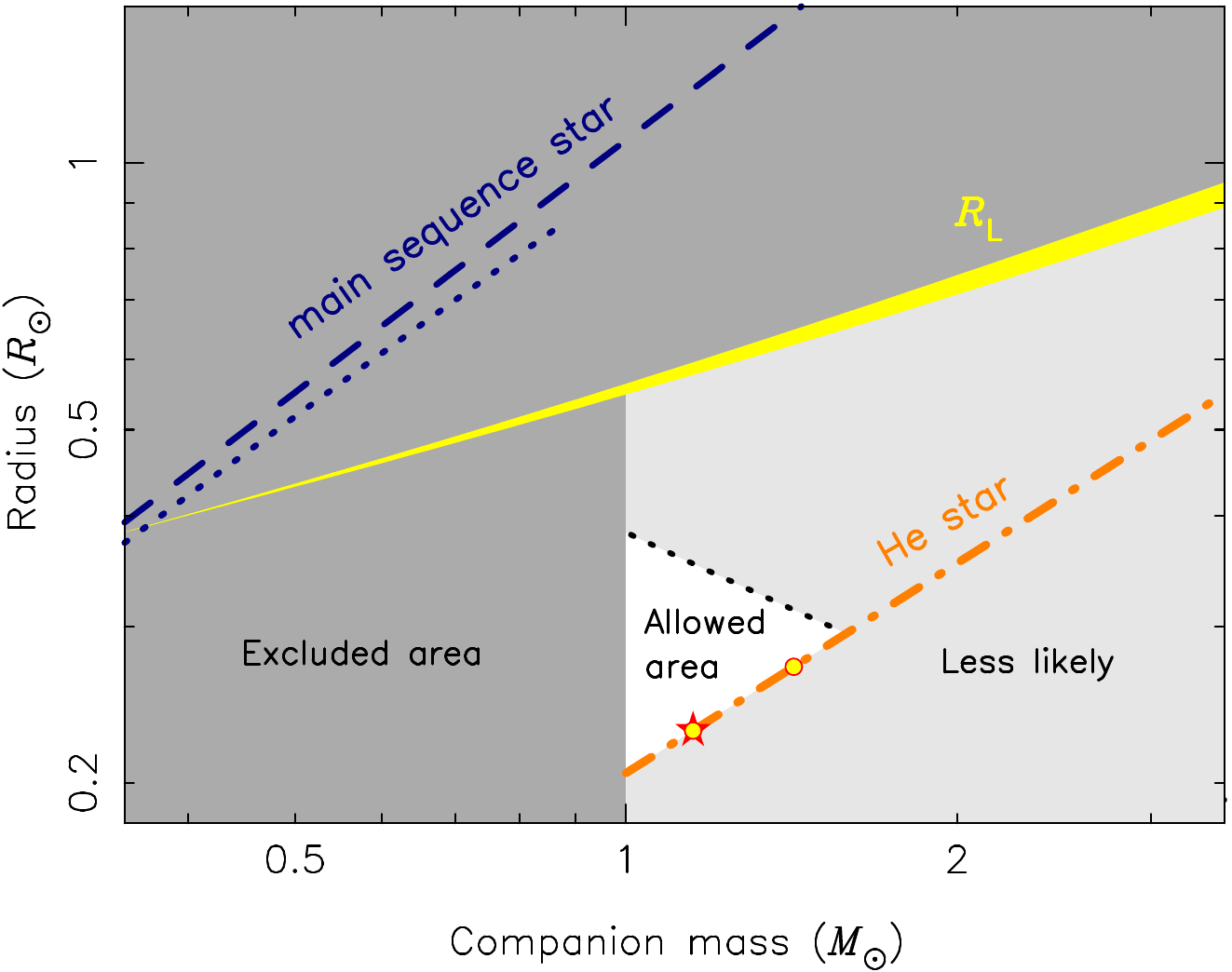}
\caption{{\bf  Parameter space for the radius and mass of the companion}. The yellow area is the Roche lobe radius R$_{\rm L}$, measured in units of solar radius ($R_\odot$), for a pulsar mass of 1.1 to 2.2~$M_\odot$ \cite{Pulsar_Astronomy2022,Tauris+2023pbse.book.....T}. The mass function (Eq.~\ref{m_f}) excludes companions $<$1.0~$M_\odot$. Dark gray shading indicates these excluded regions, which include the mass-radius relations of main sequence stars from [\cite{Demircan+1991Ap&SS.181..313D}, dashed blue line;
 \cite{Boyajian+2012ApJ...757..112B}, dotted blue line].
The orange dashed line indicates the expected mass-radius relation for helium stars with an age of 1 million years (see supplementary text). The red star and the red circle mark the radius and mass of an assumed helium star companion with $M_{\rm p}=1.4$~$M_\odot$ and $i=90^\circ$ or $i=60^\circ$ respectively. If helium stars have a hydrogen envelope they will have a large radius but regions below this line can be exclude (light gray shading).  
The black dotted line is the constraint from the non-detection at infrared wavelengths \cite{Lawrence+2007MNRAS.379.1599L}, assuming the mass-luminosity relationship of helium stars \cite{Gotberg+2023ApJ...959..125G} (see supplementary text). The parameters 
for the companion should most probably be in the allowed triangular area, less likely in the regions outside (light gray shading) though the possibility cannot be completely excluded due to uncertainties in the distance, dust extinction and common envelope evolution (see supplementary text).
}
    \label{mc-r}
\end{figure*}

The companion star cannot have a size exceeding its Roche lobe -- he gravitational
boundary at which mass would transfer from the companion to the pulsar -- otherwise the mass transfer would quench the radio pulse emission of the pulsar, contrary to our radio observations. The effective radius of the Roche lobe is as follows \cite{Eggleton+1983ApJ...268..368E}:
\begin{equation}
    R_{\rm L}=a\times\frac{0.49q^{(2/3)}}{0.6q^{(2/3)}+\ln(1+q^{(1/3)})},
	\label{rRL}
\end{equation}
where $a$ is the orbital separation of the binary stars and $q\equiv M_{\rm c}/M_{\rm p}$ is their mass ratio. Figure.~\ref{mc-r} compares this $R_{\rm L}$ constraint to possible masses and radii of the companion. A main sequence star with a mass greater than 1.0~$M_\odot$ would have a radius much larger than $R_{\rm L}$, adopting either the mass-radius relation of stars \cite{Demircan+1991Ap&SS.181..313D} or measurements from eclipsing binary systems \cite{Boyajian+2012ApJ...757..112B} (Fig.~\ref{mc-r}). Therefore, the companion cannot be a main sequence star. However, a helium star could have a radius smaller than the Roche lobe, and produce a sufficiently strong stellar wind to produce the eclipses that we observed.

We suggest that the long eclipses of PSR J1928+1815 are produced by an intra-binary bow shock  (see supplementary text) that is produced where the pulsar wind and the helium star wind collide. Because the eclipses of PSR~J1928+1815 do not exceed 50\% of the orbital phase and no irregular eclipses have been observed, we calculated that the wind from the pulsar overwhelms the helium star wind, forming a bow shock that envelops the helium star (Fig.~\ref{geometry}). Such a bow shock would absorb the pulsar radio emission in a frequency range of 1.0 to 1.5~GHz \cite{Thompson+1994ApJ...422..304T,Wadiasingh+2017ApJ...839...80W,Du+2023RAA....23l5024D}.

Alternatively, the companion of PSR~J1928+1815 could be a recently formed massive proto-white dwarf with some remnant envelope of ionized gas, which could also produce the eclipses of the pulsar. However, the lifetime of such a proto-white dwarf is three orders of magnitude shorter than that of a helium star (see supplementary text), so we consider this scenario to be unlikely. 

\paragraph*{Multiwavelength search for the companion} We performed optical imaging to search for the companion star, but no source was detected \cite{mm}. We also searched optical and infrared catalogues, but again found no corresponding source \cite{mm}. We ascribe this non-detection to the distance of PSR~J1928+1815, which is about 8~kpc \cite{mm, Cordes+2002astro.ph..7156C, Yao+2017ApJ...835...29Y}, and its location behind a spiral arm of the Milky Way, which causes a large (but undetermined) amount of extinction by the interstellar dust. Table~\ref{magnitude} lists the observed limits at different wavelengths, compared to the estimated brightness of a helium companion star with various masses located at that distance and the estimated extinctions. The tightest constraint comes from the non-detection in the $K$-band  \cite{mm,Lawrence+2007MNRAS.379.1599L}
which indicates that the companion mass is probably less than 1.6~$M_\odot$. The optical non-detections are less constraining, owing to the higher dust extinction at optical wavelengths.

Binary pulsar systems sometimes produce high energy emission \cite{Roberts+2013IAUS..291..127R,Koljonen+2023MNRAS.525.3963K}. We searched available x-ray and $\gamma$-ray catalogs  \cite{Boller+2016A&A...588A.103B,Evans+2020ApJS..247...54E,Abdollahi+2022ApJS..260...53A}, but did not find any possible counterpart of PSR J1928+1815, probably because of too few high-energy photos reaching us from such a large distance.

\paragraph*{Binary stellar evolution model} 

We investigated evolutionary scenarios that could produce a binary system like PSR~J1928+1815. The short spin period of PSR~J1928+1815 indicates that the system has recently experienced hypercritical mass accretion process during a common envelope phase (see supplementary text). This process might also have occurred in intermediate-mass binary pulsars that has a white dwarf companion with a mass around 1~$M_\odot$, such as PSR J1952+2630 \cite{Lazarus+2014MNRAS.437.1485L}. 

\begin{figure}
\centering
    \includegraphics[width=0.55\columnwidth]{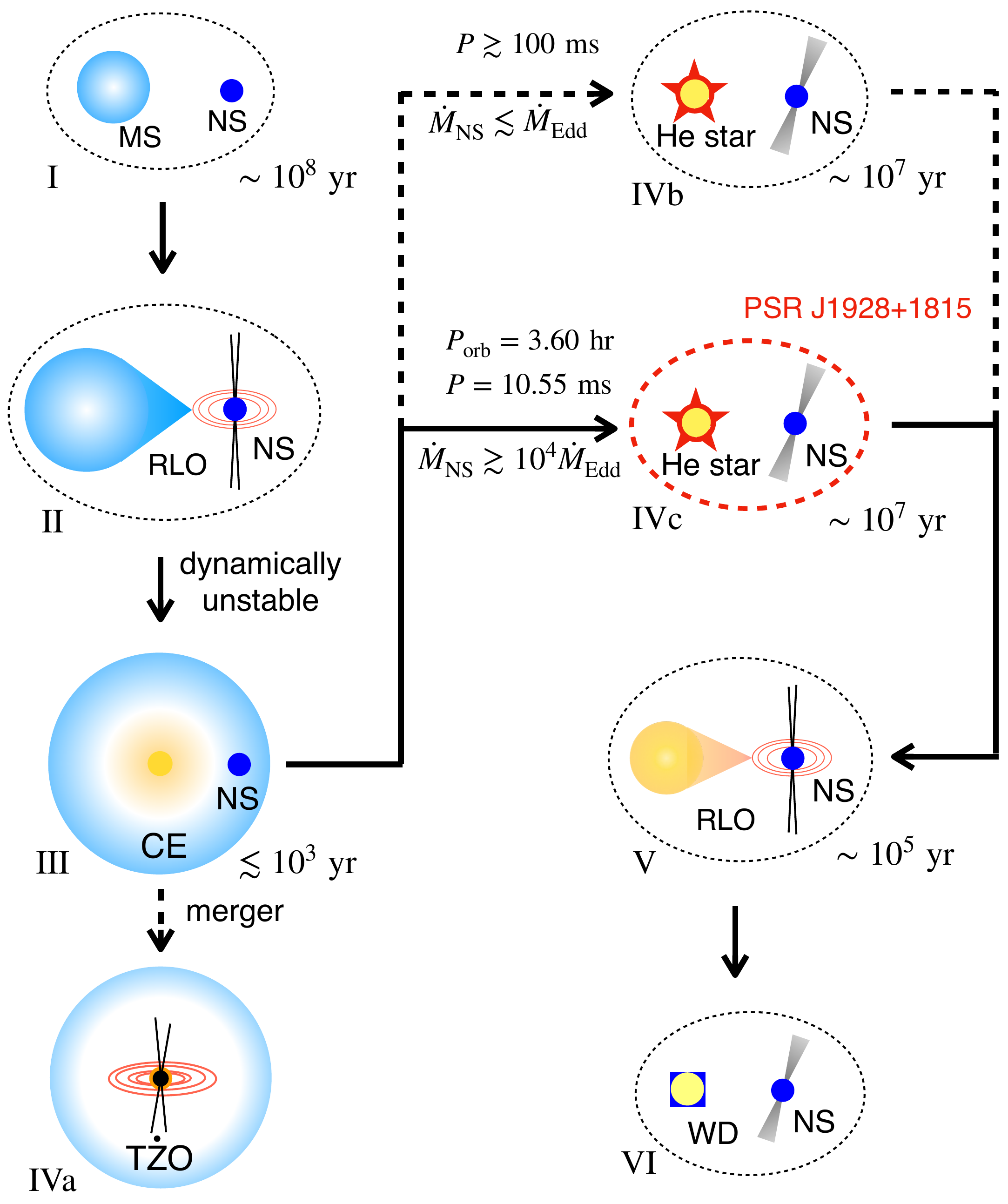}
   \caption{{\bf Proposed stellar evolution of the PSR~J1928+1815 binary system.} Stages are labelled with Roman numerals, connected by arrows. Stage I consists of a main sequence (MS) star and a neutron star (NS). Once the MS has exhausted the hydrogen in its core, it  expands until it overfills its Roche lobe, causing mass transfer via the Roche lobe overflow (RLO) in stage II, and its mass loss causes the further expansion so that the binary enters the common envelope (CE) phase at stage III \cite{Ivanova+2013A&ARv..21...59I}.  During the CE phase, the orbit shrinks rapidly, for about $\sim10^3~\rm yr$, with three possible outcomes.  If the orbit shrinks more rapidly than the envelope is ejected, the NS merges with the core of the MS to produce a Thorne-\.Zytkow object (T\.ZO) \cite{Thorne+1975ApJ...199L..19T,Thorne+1977ApJ...212..832T} (stage IVa). If the envelope is ejected before the merger occurs, the stripped core of the MS becomes a helium star in a compact orbit with the NS (stages IVb for photon cooling and IVc for neutrino cooling \cite{Houck+1991ApJ...376..234H}, see supplementary text). The short spin period of PSR J1928+1815 (red dashed ellipse) indicates that it  accreted at least 0.01~$M_\odot$ probably during the common envelope phase, at an accretion rate $\dot{M}_{\rm NS}$ higher than $\sim10^4\dot{M}_{\rm Edd}$ \cite{Houck+1991ApJ...376..234H}. 
   The helium star is probably not massive enough to produce a supernova, so we predict it will probably expend to cause a second RLO at stage V, followed by the formation of a neutron star-white dwarf (NS-WD) binary at stage VI.
   }
   \label{evolution}
\end{figure}

We interpret PSR~J1928+1815 as a binary system consisting of a helium star 
and a radio pulsar that recently experienced a common envelope phase (see supplementary text). Our proposed evolutionary scenario is illustrated in Fig.~\ref{evolution}. Although it is unclear whether a millisecond pulsar can be formed directly from a supernova \cite{Tauris+2013A&A...558A..39T}, they are known to be formed by mass transfer from a binary companion to a neutron star, because the accreting mass increases the spin rate \cite{Alpar+1982Natur.300..728A,Radhakrishnan+1982CSci...51.1096R}. 
If the companion is more massive than the accreting neutron star, the mass-loss rate of the companion can substantially exceeds the mass accretion rate of the neutron star, owing to dynamical instability in the mass transfer process \cite{Ge+2015ApJ...812...40G,Ge+2020ApJ...899..132G}, which causes a common envelope to form (Fig.~\ref{evolution}) \cite{Ivanova+2013A&ARv..21...59I}. 
During the common envelope stage the system's orbital momentum and energy are partially transferred into the envelope, causing the orbit to shrink rapidly. The outcome of this spiral-in phase is either (i) a merger of neutron star and the core of the companion, producing a massive neutron star or a black hole, or (ii) ejection of the common envelope, leaving a very compact binary system \cite{Paczynski+1976IAUS...73...75P,Iben+1993PASP..105.1373I,Ivanova+2013A&ARv..21...59I}. Many observed compact neutron star binaries have been interpreted as having formed through this common envelope evolution channel, including the double neutron stars and compact binary systems consisting of a pulsar and a massive white dwarf with an orbital period less than 3 days \cite{Tauris+2023pbse.book.....T}. 

This scenario can form a helium star companion in an orbital period of several hours \cite{Tauris+2011MNRAS.416.2130T,Ivanova+2013A&ARv..21...59I}, consistent with the 3.6-hour orbit we measured for the PSR J1928+1815 system. Ejection of the envelope would leaves a helium star, in the core helium-burning phase, if the donor star was on the red giant branch (RGB) at the beginning of the common envelope phase \cite{Dewi+2002MNRAS.331.1027D}. If the donor was instead an asymptotic giant branch (AGB) star at the beginning of the common envelope phase, the ejection would leave  a helium star with a carbon-oxygen core \cite{Dewi+2002MNRAS.331.1027D}. We favor the former case for the PSR J1928+1815 binary, because it produces a radio pulsar-helium star system with a much longer lifetime. Assuming $M_{\rm p}=1.4~M_\odot$ and $\sin i=1$, we calculated that the orbital energy of PSR~J1928+1815 is about $-2\times10^{48}~\rm erg$. An RGB progenitor of a $1\text{-}M_\odot$ helium star has a binding energy of ($1.3~\rm to~2.2)\times10^{48}~\rm erg$ \cite{Tauris+2011MNRAS.416.2130T}. Therefore. the orbital energy in PSR~J1928+1815 is sufficient to be the main energy source that dispersed the envelope of an RGB progenitor \cite{Ivanova+2013A&ARv..21...59I}. Our simulations suggest that the RGB progenitor of the companion had an initial mass of 5 to 8~$M_\odot$ (see supplementary text and fig.~\ref{ge-cee}). 

For the pulsar to reach a 10.55 ms spin period, we calculated that at least 0.01~$M_\odot$ of the envelope must have been accreted to the neutron star during the mass transfer phase (Fig.~\ref{evolution}). For spherical accretion, if gravitational potential energy is full converted entirely into radiation, the maximum possible mass accretion rate is called the Eddington accretion rate $\dot{M}_{\rm Edd}$. The common envelope stage lasts for $<$1000 years, and the Eddington accretion rate of a neutron star accreting hydrogen-rich material is only $\sim2\times$10$^{-8}$~$M_\odot$\,year$^{-1}$, therefore, an over-critical accretion is required during which the gravitational potential energy is carried away by neutrinos instead of photons so the mass accretion rate can be much higher than $\dot{M}_{\rm Edd}$ \cite{Houck+1991ApJ...376..234H}. 

We simulated the evolution of a synthetic population of binary stars(see supplementary text). We ound that the formation rate of millisecond pulsar binaries with a helium-star companion of  $1.0$~ to~$2.0~M_\odot$ and an orbital period of $0.1~\rm to~0.2$ days is $(1.3~\rm to~7.2)\times10^{-6}~ year^{-1}$, depending on the assumptions made (table~\ref{birthrate and number}). In the Milky Way, we expect 16~to~84 such systems (see supplementary text). Not all of these are potentially detectable, because pulsar emission occurs in a narrow beam and distant sources would be too faint to detect.

The companion helium star in this binary has a mass of 1.0 to 1.6~$M_\odot$, too low to produce a supernova. We expect it to evolve for about 10 million years from its formation and then experience matter transfer (
Fig.~\ref{evolution}) \cite{Tauris+2023pbse.book.....T}. This mass transfer phase last $\sim10^5$ year, after which the system would become a detached binary millisecond pulsar with a massive white dwarf companion in a slightly wider orbit (Fig.~\ref{evolution}). Our simulations (see supplementary text) predict that the binary will then evolve to a system consisting of a neutron star with a mass of 1.41~$M_\odot$ and a carbon-oxygen white dwarf companion with a mass of 0.82~$M_\odot$, in a compact orbit with a period of 6.2 hours  (see supplementary text and fig.~\ref{casebb_yl}). 

\paragraph*{Spin-orbit misalignment}
As discussed in the previous sections, we interpreted this binary system as having formed from a common envelope stage, which must have been preceded by a mass accretion stage. The angular momentum from the accreting material spin up the neutron star, causing 
the spin axis of the neutron star to align with the orbital angular momentum of the binary\cite{Bhattacharya+1991PhR...203....1B}. The observed radio pulsar eclipses of PSR~J1928+1815 indicate the binary has a highly inclined, nearly edge-on orbit, so we expect the pulsar spin axis to be roughly perpendicular to our line of sight. 

\begin{figure}
\centering
    \includegraphics[width=0.7\columnwidth]{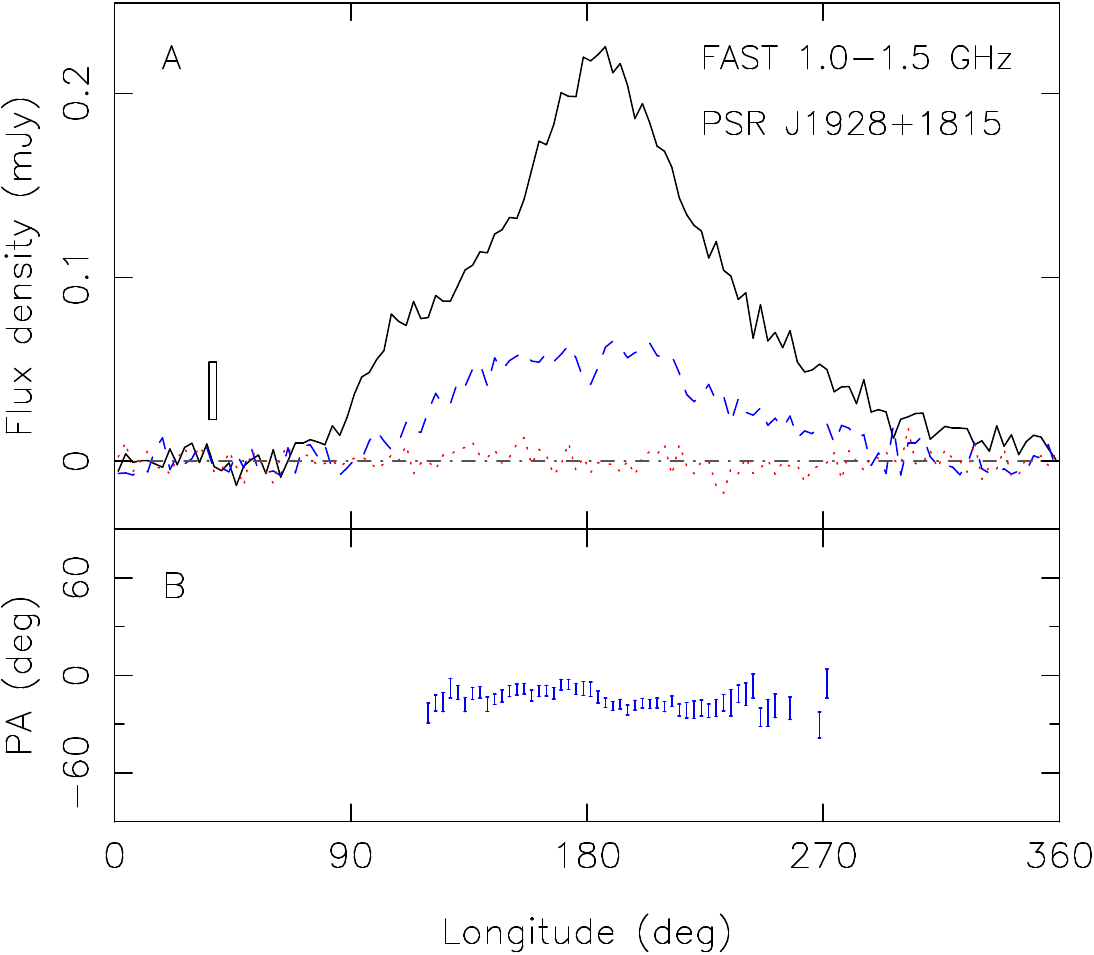}
   \caption{{\bf Mean polarized pulse profiles of PSR~J1928+1815.} The pulse profiles are obtained from the sum of all available FAST tracking observations in the frequency range of 1.0 to 1.5 GHz with a total integration time of 14.5~hr \cite{mm}. Radio frequency interference and eclipses were manually removed. ({\bf A)} The total intensity profile (solid black line), linear polarized profile (dashed blue line), and circular  polarized profile (dotted red line, where positive for the sense of left-hand circular polarization) are plotted as a function of pulsar rotation longitude, with a scale-box indicating the 1-bin width and the $2\sigma$ of flux density. ({\bf B}) The polarization angles with $\pm1\sigma$ error-bar are plotted for the phase-bins with a linearly polarized intensity $>3\sigma$.} 
   \label{Poln-prof} 
\end{figure}

The  pulse profile of PSR~J1928+1815 (Fig.~\ref{Poln-prof}, achieved by 14.5-hour observations by FAST) is very wide, occupying more than 300$^\circ$ in the pulsar rotation phase, with an almost constant polarization angle. 
For normal pulsars, the radio emission is produced in the open magnetic field region near the magnetic poles around the magnetic axis. 
For millisecond pulsars, the emission beam is generally wider and more complex than that for normal pulsars, which has previously been attributed to the multi-polar magnetic fields \cite{Krolik+1991ApJ...373L..69K}. The wide pulse profile of PSR~J1928+1815 indicates that our line of sight intersects the emission beam for more than two-thirds of 
the pulsar rotation. According to the rotating vector model \cite{Radhakrishnan+1969ApL.....3..225R} of pulsar emission, this implies that the angles between the spin axis, the magnetic axis, and the line of sight are small or that the emission beam is very wide. The constant polarization angle indicates that there is a simple emission geometry, without any multi-polar fields near the magnetic poles involved for emission. 
This raises a contradiction: Such a radio emission geometry would not be visible from Earth for an edge-on binary orbit because the spin and magnetic axes of the neutron star are both perpendicular to the line of sight, 
contrary to the general theoretical expectation for accretion spinning-up of the neutron star.

A theoretical study of evolution of the spin and the magnetic axes of accreting neutron stars \cite{Biryukov+2021MNRAS.505.1775B}
has demonstrated that a rigid spherical accretion neutron star in a wind-fed binary gains a random angular momentum from accreted matter, causing a non-linear random-walk evolution of the spin and magnetic axes, and finally
the two axes could be either aligned or perpendicular. 
We suggest this scenario can apply to the embedded neutron star in the common envelope phase and then explain the wide polarization pulse profile of PSR~J1928+1815 and its emission geometry.


\baselineskip18pt

\clearpage
\newpage

\clearpage
\newpage


\begin{table*}
     \centering
     \caption{{\bf Parameters of the PSR~J1928+1815 binary system derived from 
     the FAST observations}. Date and time were expressed in the modified Julian date (MJD). The TOAs were fitted with a binary pulsar timing model \cite{Hobbs+2006MNRAS.369..655H, mm}. The uncertainty of the model parameters  is the 68.3$\%$ confidence level (equivalent to 1$\sigma$ uncertainty). 
     }
    \vspace{5mm}
    \label{timing solution}
    \setlength{\tabcolsep}{1mm}{
    \renewcommand\arraystretch{0.9}
    {
    	\begin{tabular}{lc} 
		\hline
		Parameter      & Value\\
		\hline                                 
Observation date span (MJD)         &   58999 to 60668\\
Number of TOAs      &                 152\\
Reduced $\chi^2$ value &  1.296 \\
MJD of period determination  &        59408\\
\hline
Right ascension, $\alpha$ (J2000 equinox)   &        19$^{\rm h}$28$^{\rm m}$08$^{\rm s}$.34901$\pm0.00041$\\
Declination, $\delta$ (J2000 equinox)       &         +18$^\circ$15$'$30$''$.270$\pm0.010$\\
Spin period, $P$               &        0.0105494989888508$\pm0.0000000000000078$ s\\
Spin period derivative, $\dot{P}$  &     (3.63405$\pm0.00022$)$\times10^{-18}$ s\,s$^{-1}$\\              
Spin frequency, $\nu$  &       94.791231418368$\pm0.000000000070$ Hz\\
Spin frequency derivative, $\dot{\nu
}$ &  $(-3.26534\pm0.00020)\times10^{-14}$ Hz\,s$^{-1}$\\
Spin-down luminosity, $\overset{.}{E}$ & 1.2$\times10^{35}$ erg\,s$^{-1}$\\
Characteristic age,  $\tau_c$       &          46 million years\\
Dispersion measure, DM       &  346.158$\pm0.014$ pc\,$\rm cm^{-3}$\\
Faraday rotation measure, RM & 254.3$\pm3.0$ rad\,$\rm m^{-2}$\\
\hline
Orbital period, $ P_{\rm orb}$           &       0.14990246322$\pm0.00000000028$ days\\
Rate of change of orbital period, $\dot{P}_{\rm orb}$   &  $(-4.3\pm4.5)\times10^{-13}$ s\,s$^{-1}$\\
Projected semi-major axis, $x$ & 1.698628$\pm0.000020$ light-seconds\\
Rate of change of $x$, $\dot{x}$  & $(1.9\pm3.8)\times10^{-13}$ light-seconds\,s$^{-1}$\\
Time of ascending node passage, $ T_{\rm asc}$ (MJD)  &  59211.18761057$\pm0.00000024$\\
First Laplace parameter, $\epsilon_1=e\sin\omega$             &          (-0.4$\pm$1.8)$\times10^{-5}$\\
Second Laplace parameter,  $\epsilon_2=e\cos\omega$             &           (0.7$\pm$1.5)$\times10^{-5}$\\
Mass function       &    0.2341854$\pm0.0000074$ $M_\odot$  \\
Coalescence time, $ \tau_{\rm GW}$  &    0.26 billion years\\
		\hline
    \end{tabular}
    }
    }
\end{table*}

\clearpage
\bibliographystyle{sciencemag}

\newpage

\section*{Acknowledgments}
We thank S. Ransom and two other anonymous referees for their careful reading and thoughtful suggestions. We thank B. Q. Chen for helpful discussions on the extinction model of the Galactic disk. This work made use of data from FAST (https://cstr.cn/31116.02.FAST), which is a Chinese national megascience facility, built and operated by the National Astronomical Observatories, Chinese Academy of Sciences. This work is based in part on data obtained as part of the UKIRT Infrared Deep Sky Survey. UKIRT is operated by the Joint Astronomy Centre on behalf of the UK Particle Physics and Astronomy Research Council. The Pan-STARRS1 Surveys (PS1) and the PS1 public science archive have been made possible through contributions by the Institute for Astronomy, University of Hawaii; the Pan-STARRS Project Office; the Max-Planck Society and its participating institutes; the Max Planck Institute for Astronomy, Heidelberg; the Max Planck Institute for Extraterrestrial Physics, Garching; The Johns Hopkins University; Durham University; the University of Edinburgh; the Queen’s University Belfast; the Harvard-Smithsonian Center for Astrophysics; the Las Cumbres Observatory Global Telescope Network Incorporated; the National Central University of Taiwan; the Space Telescope Science Institute; NASA under grant no. NNX08AR22G issued through the Planetary Science Division of the NASA Science Mission Directorate; National Science Foundation grant no. AST-1238877; the University of Maryland; Eotvos Lorand University (ELTE); the Los Alamos National Laboratory; and the Gordon and Betty Moore Foundation. 
\paragraph*{Funding:} 
The authors were supported by the Natural Science Foundation of China (NSFC), grant numbers 11988101 (ZLY, JLH, DJZ, WCJ, TW, SW, JFL, WQS, LGH, CW, PFW, JX, and YY), 11833009 (JLH and JX), 12133004 (HGW, PFW, CW and XPY), 12225303 (PJ), 12225304 (BW), 12273057 (SW), 12273014 (WCC), 12041304 (NW), and 12288102 (NW). JLH and JX were supported by the  National SKA Program of China 2020SKA0120100, and JLH, CW, PFW and JX) were also supported by the Chinese Academy of Sciences  a project via project JZHKYPT-2021-06.
%
HWG acknowledges support from the National Key R\&D Program of China (grant no. 2021YFA1600403), NSFC (grant Nos. 12288102, 12173081), the key research program of frontier sciences, CAS (grant no. ZDBS-LY-7005), and Yunnan Fundamental Research Projects (grant no. 202101AV070001). 
XDL was  supported by the National Key Research and Development Program of China (2021YFA0718500) and NSFC (grant nos. 12041301 and 12121003). 
BW was  supported by the Western Light Project of CAS (no. XBZG-ZDSYS-202117).
%
\paragraph*{Author contributions:}
JLH detected the pulsar, coordinated and supervised all follow-up investigations, and led the manuscript writing. 
DJZ identified this pulsar as being in a binary system and obtained the initial orbit period. 
WCJ used the eclipses to constrain the binary companion.
ZLY carried out follow-up FAST observations, processed all FAST data and fitted the pulsar timing model, determined the eclipse phase range and the companion star, and helped drafted the paper.
TW developed the procedure for FAST data preparation  and carried out  follow-up observations.
WQS obtained the acceleration parameter of this pulsar, and also constrained the timing difference between FAST beams. 
WCC interpreted the  system, and constrained  companion star brightness. 
HWG simulated the common envelope evolution and potential progenitors of the system.
WCC and YS performed population synthesis simulations.
LGH, SW, and JFL investigated the optical image of the companion and searched optical and infrared catalogues.
XDL and WCC investigated the binary evolution of this system and emphasized all possible evolution channels.  
LHL, YLG, and BW simulated the future evolution of this binary and investigated potential companions.
YS also modeled the common envelope evolution. 
PFW developed the procedures for FAST pulsar polarization and timing processing. 
CW scheduled the FAST observation.
YY determined the astrometric of this pulsar. 
JX maintained the FAST data-processing platforms.
JFL and JLH discussed the polarization profile and pulsar emission geometry. 
ZSY searched catalogues for x-ray and $\gamma$-ray emission of this binary. 
JLH, CW, PFW, HGW, XPY, RXX, NW, DJZ, TW, WCJ, SQW, HX, JX, WJH, BJW, WQS, JBW, WMY, CFZ, PJ, JHS, JCJ and DJY initially proposed the FAST key science project for the GPPS survey, and PJ and JHS ensured FAST observation data quality used in this paper.
All authors discussed the results and commented on the manuscript. 
\paragraph*{Competing interests:}
The authors declare no competing interests.
\paragraph*{Data and materials availability:}
 
All FAST data of PSR~J1928+1815 published in this paper are available on \url{http://zmtt.bao.ac.cn/GPPS/gpps0121}.

\subsection*{Supplementary Materials:}
Materials and Methods\\ 
supplementary text\\
Figs. S1 to S8, 
Tables S1 to S3\\
References (46–111)\\


\clearpage
\newpage

\renewcommand{\thefigure}{S\arabic{figure}}
\renewcommand{\thetable}{S\arabic{table}}
\setcounter{figure}{0}
\setcounter{table}{0}
\renewcommand{\theequation}{S\arabic{equation}}
\renewcommand{\thepage}{S\arabic{page}}
\setcounter{equation}{0}
\setcounter{page}{1} 

\begin{center}
\section*{Supplementary Materials for\\ \scititle}

\author{
Z. L. Yang,
J. L. Han$^{\ast}$, 
D. J. Zhou,  
W. C. Jing,
W. C. Chen,  
Tao Wang,  
X. D. Li,
Song Wang, 
Bo Wang, 
H. W. Ge, 
Y. L. Guo, 
L. H. Li,  
Yong Shao,
J. F. Liu,
W.~Q. Su
L. G. Hou, 
W. J. Huang,
J. C. Jiang, 
Peng Jiang, 
J. H. Sun,
B. J. Wang, 
Chen Wang,
H. G. Wang, 
J. B. Wang,  
Na Wang,  
P. F. Wang, 
S. Q. Wang, 
Heng Xu,  
Jun Xu, 
R. X. Xu,  
W. M. Yan, 
Yi Yan, 
X. P. You,
D. J. Yu,
Z. S. Yuan, and 
C. F. Zhang
\\[2mm]
%
$^\ast$Corresponding author. Email: hjl@nao.cas.cn\\
}
\end{center}

\subsubsection*{This PDF file includes:}
Materials and Methods\\
supplementary text\\
Figures S1 to S8\\
Tables S1 to S3

\newpage

\subsection*{Materials and Methods}\label{sec3}

\subsubsection*{FAST observations of PSR~J1928+1815}\label{FASTobs}

PSR~J1928+1815 was identified during the FAST GPPS survey in a snapshot observation on 2020 May 31. It was later confirmed by a verification tracking observation for 15 minutes on 2020 August 10. Phase shifts were evident in both observations, indicating its binary nature. It has high orbital acceleration (73.0~$\rm m~s^{-2}$) \cite{Han+2021RAA....21..107H}. 25 follow-up tracking observations have been made, as listed in Table~\ref{obsinfo}. All FAST observations were made using the L-band 19-beam receiver \cite{Jiang+2020RAA....20...64J} working in the radio frequency range of 1.0 to 1.5~GHz with 2048 or 4096 frequency channels. In most observations, the data for four 4 polarization channels, obtained from $X$ and $Y$ signals from the linear polarization feed, $XX$, $YY$, Re[$X^{*}Y$] and Im[$X^{*}Y$] were recorded with a sampling rate of 49.152 \textmu s for every channel. Modulated 1.1~K noise signals with a period of 2.01326~s were injected into the receiver feeds for 2~minutes before or after the observation sessions, which were used to calibrate polarization data.  

Among those observations, three longer tracking observations were carried out on 2012 March 13, 2021 April 5, and 2023 December 18, which covered large parts of the orbit. These showed the long eclipses in the orbital phase range of 0.18 to 0.35 ( Fig.~\ref{Timing result & flux vs. orbital phase} and Fig.~\ref{time-phase}). The first two observations were performed with an offset of 1.1 arcminutes between the pulsar and the telescope beam center, so their measured flux densities were corrected by a gain factor of 4/3 in Figs.~\ref{Timing result & flux vs. orbital phase},~\ref{Poln-prof}~and~\ref{time-phase}. PSR~J1928+1815 was not detected in three short (15- or 20-minute) observing sessions because they fell in the eclipse phase range (Table~\ref{obsinfo}).

\subsubsection*{Pulsar timing and polarized pulse profiles}\label{timing}

Barycentric spin periods $P_{\rm bary}$ and accelerations were determined using the package \textsc{Presto} (package version: 2020 October 25) \cite{Ransom+2001PhDT.......123R}. These values from observation sessions in 2020 were plotted in two dimensions (Fig.~\ref{P0-a}); they form an ellipse from which the orbital period and the projected semi-major axis were obtained \cite{Freire+2001MNRAS.322..885F}, and the first rough estimate of the orbital period is 3.6~hours. The orbital parameters were refined as more observations were collected, using the method of \cite{Bhattacharyya+2008MNRAS.387..273B}. After correcting the phase shifts within an observation session due to orbital motions, we fold observation data into 128-bin pulse profiles using \textsc{Dspsr} \cite{Straten+2011PASA...28....1V}. 

For timing analysis, we extracted TOAs from observation data using the package \textsc{Psrchive} (package version: 2022 January 14) \cite{Hotan+2004PASA...21..302H}. After polarization calibration with the command \texttt{pac} in the \textsc{Psrchive} package, we remove radio frequency interference (RFI) automatically using the  method given in the reference \cite{Chen+2023RAA....23j4004C}, and checked the results manually using \texttt{psrzap}, then integrate all frequency channels to produce the total intensity profiles using the command \texttt{pam}. The profiles are then compared to a template produced by the command \texttt{paas}, and then TOAs were extracted from all observations using the command \texttt{pat}. Excluding these values in the eclipse range of 0.18 to 0.35 orbital phase, we were left with 152 TOAs, one from each 5-minute segment of FAST observations, spanning 4.5 years (Table~\ref{obsinfo}). 

Using the pulsar timing analysis package \textsc{Tempo2} (package version: 2023 May 1) \cite{Hobbs+2006MNRAS.369..655H}, we determined a phase-coherent timing solution for PSR~J1928+1815 from these TOAs (Table~\ref{timing solution}). The number of freedom is 140 and the reduced $\chi^2$ is 1.296. Fig.~\ref{Timing result & flux vs. orbital phase} shows TOA residuals as a function of epoch and orbital phase. The Solar System ephemeris DE440 \cite{DE440} was applied to correct the motion of the FAST relative to the barycenter of the Solar System, while the motion of the pulsar was described using the ELL1 model \cite{Lange+2001MNRAS.326..274L} -- a commonly used model for low eccentricity orbits of binary stars, in which three Keplerian orbital parameters, the time of periastron passage $T_0$, eccentricity $e$ and the longitude of periastron $\omega$, are replaced by the time of ascending node passage $T_{\rm asc}$ and two Laplace parameters $\epsilon_1\equiv e\sin\omega$ and $\epsilon_2\equiv e\cos\omega$. 
The DM value of PSR~J1928+1815 is refined by comparing the accumulated sub-band profiles from all FAST observations, and then is applied for the mean pulse profiles and timing analysis.

Using the phase-coherent timing solution, all data were refolded using the revised pulsar ephemeris, and all tracking observations with polarization data were summed together to determine the mean pulse polarization profile over the rotation longitude of a spin period (Fig.~\ref{Poln-prof}). From the polarization data of the reduced frequency channels in the FAST observation band, we get the Faraday rotation measure of $254.3\pm0.8~\rm rad~m^{-2}$ (68\% confidence level) for PSR~J1928+1815. 

PSR~J1928+1815 has a period of 10.55~ms and a high period derivative of $3.6\times10^{-18}$ s\,s$^{-1}$. These values are similar to those of the binary PSR~J1952+2630 (20.7~ms and  $4.2\times10^{-18}$ s\,s$^{-1}$) \cite{g2022J1952}. Both systems have show a similarly young characteristic ages, but PSR J1952+2630 has no eclipses.

\subsubsection*{Searching for the companion in optical and infrared bands}\label{optobs}

The helium star companion of PSR~J1928+1815 was constrained above to have a mass of 1~to~2~$M_\odot$. We searched for, but did not detect, any counterparts in optical and infrared bands.  
The distance of this binary system was estimated from the pulsar DM value, using two Galactic electron density distribution models \cite{Cordes+2002astro.ph..7156C, Yao+2017ApJ...835...29Y}. We find values of 9.7 or 7.2~kpc from the two models, therefore we assume a distance of 8~kpc for following estimations.

Before the exact timing solution for PSR~J1928+1815 was obtained, we searched for an optical companion using the coarse position of the pulsar. The optical imaging observations were carried out for 2.5~hr on 2021 May 20,  using the 85-cm telescope at the Xinglong station of the National Astronomical Observatories of China (NAOC). The $5\sigma$ limiting magnitudes for the 300-s exposures were 21.1, 20.6 and 20.3~mag in the $V$, $R$ and $I$ bands respectively. We performed differential photometry (with nearby bright non-variable stars) for all objects within the position uncertainty of 1.5 arcminutes, and did not find  possible optical counterpart with a periodic signal in the light curves. After additional pulsar observations were obtained, the precision of its position is improved to 0.01 arcseconds, at which point we re-analyzed the optical images. No  optical counterpart can be found at the position in the images. 

We then searched for an optical or infrared counterpart from public optical archive data at the precise position of the pulsar using the VizieR catalog access tool \cite{vizier}. No object was found within 3 arcsec from the available survey catalogs, including the Two Micron All Sky Survey (2MASS) \cite{Skrutskie+2006AJ....131.1163S}, the Galactic Legacy Infrared Midplane Survey Extraordinaire (GLIMPSE) \cite{Benjamin+2003PASP..115..953B}, 
the Gaia Early Data Release 3 (Gaia DR3) \cite{Gaia+2021A&A...649A...1G}, and combined catalog of Wide-field Infrared Survey Explorer (ALLWISE) \cite{Wright+2010AJ....140.1868W,Mainzer+2011ApJ...731...53M}.
We also analyzed an archival optical image  (Fig.~\ref{pans-ukirt}) from the 3$\pi$ Steradian Survey in the first part of Panoramic Survey Telescope and Rapid Response System (Pan-STARRS1) that was observed by using a 1.8-meter telescope on Haleakala in Hawaii \cite{Chambers+2016arXiv161205560C}. In the $g, r, i, z, y$ bands, the mean 5-sigma point source sensitivities are 23.3, 23.2, 23.1, 22.3, 21.4~mag, respectively. No optical counterpart was found within a radius of 2~arcseconds from PSR~J1928+1815. We therefore conclude that the companion should is fainter than a magnitude of 23~mag in the $gri$ bands.

We also searched for the infrared counterpart from the archive data of the United Kingdom Infrared Telescope (UKIRT) Infrared Deep Sky Survey (UKIDSS)  \cite{Lawrence+2007MNRAS.379.1599L}. UKIDSS uses the UKIRT Wide Field Camera \cite{2007A&A...467..777C}, photometric system \cite{2006MNRAS.367..454H}, and a pipeline processing and science archive \cite{Hambly+2008MNRAS.384..637H}. We downloaded the archival data from the Wide Field Camera (WFCAM) Science Archive \cite{Hambly+2008MNRAS.384..637H}. The Galactic Plane Survey (GPS) within UKIDSS reached limiting sensitivities of 19.9, 19.0, and 18.8~mag in the $JHK$~bands respectively \cite{Lawrence+2007MNRAS.379.1599L}. There is a faint neighbor source about 1~arcsec away from the radio position (Fig.~\ref{pans-ukirt}), 
with the  magnitudes of $J=$18.8$\pm$2.0, $H=$19.1$\pm$0.5, and $K=$18.5$\pm$1.1~mag. Considering the large offset of 1~arcsec from the timing position, it is unlikely to be the counterpart of PSR~J1928+1815's companion. We therefore conclude that the companion is fainter than 19.9, 19.0 and 18.8~mag at the $JHK$-bands. 

\subsection*{supplementary text}

\subsubsection*{Intrabinary bow shock and eclipses }

Following the interpretation of other eclipsing binary pulsars (Fig.~\ref{Pb-m2}), we attribute the regular pulsar eclipses of PSR~J1928+1815 to intrabinary material \cite{1988Natur.333..237F}, forming a stable bow shock (enveloping its companion) where stronger pulsar winds and weaker helium star winds are balanced \cite{Thompson+1994ApJ...422..304T,Wadiasingh+2017ApJ...839...80W}.  If the companion helium star of PSR~J1928+1815 has a mass greater than 2~$M_\odot$, stronger stellar winds would overwhelm the pulsar winds, producing much longer eclipses than we observed or causing a large mass flow to the pulsar that would quench radio emission of the pulsar. Therefore, the mass range of the helium star companion is 1 to 2~$M_\odot$ from the requirements of wind momentum and pulsar emission.  Observations of stripped helium stars with a mass of less than 4~$M_\odot$ \cite{Drout+2023Sci...382.1287D,Gotberg+2023ApJ...959..125G} have shown a much smaller wind mass loss rate than was expected theoretically, so that companion helium star could have a slightly higher mass. We further constrain the companion mass using optical/infrared observations below.

The momentum flux of pulsar wind can be estimated using pulsar spin-down luminosity $\dot{E}$ via $\dot{E}/c=4\times10^{19}~\rm kg\,m\,s^{-2}$, here $c$ is the speed of light.  Assuming the solar metallicity $Z=0.02$ -- the mass fraction of all elements heavier than helium -- for the helium star companion with a mass of 1.1~$M_\odot$, according to theoretical predictions \cite{Vink+2017A&A...607L...8V,Gotberg+2018A&A...615A..78G} the wind mass-loss rate should be about $\dot{M}=10^{-10}~M_\odot~\rm yr^{-1}$ and the wind terminal velocity is about $2\times10^3~\rm km~s^{-1}$. Therefore the total momentum flux of the companion accumulated over the 4$\pi$ solid angle is $\dot{M} v\sim1\times10^{19}~\rm kg~m~s^{-2}$, where $v$ is the velocity of the helium star wind. The wind mass loss rate generally increases with the companion mass and metallicity. 

The ratio of the two wind momentum fluxes, $\eta_{\rm w} \equiv \dot{M}vc/\dot{E}$, determines the geometry of the bow shock, which is closely related to the eclipse range. The inferred geometry of the pulsar and companion is illustrated in Fig.~\ref{geometry}. We define the eclipse radius as being the distance $d$ from the companion star center to the connection line between the pulsar and the observer at the ingress phase of $\phi_{\rm in}$ or the egress phase of $\phi_{\rm eg}$, along the circular orbit of the companion.  The distance $d$ is related to the orbital phase $\phi$
and the binary separation $a$ by 
$[a\sin(2\pi\phi)\sin i]^2= a^2-d^2.$ 
If the eclipses extend to over a half of the orbit it will be hard to define an eclipse radius. Assuming a pulsar mass of a 1.4~$M_\odot$ and $\sin i=1$, we estimate the minimum separation $a=1.62~R_\odot$. The two eclipse radii from the companion center are $0.69~R_\odot$ and $0.95~R_\odot$ at the orbit phases at $\phi_{\rm in}=0.18$ and $\phi_{\rm eg}=0.35$, respectively, larger than the Roche lobe radius $R_{\rm L}=0.58~R_\odot$ of the companion calculated using Eq.~\ref{rRL}. Because the real inclination angle $i$ must be $<90^{\circ}$, these values therefore are the lower limits.
The shortest distance $R_0$ of the intrabinary bow shock from the helium star depends on the ratio of two momentum fluxes, $\eta_{\rm w}$, via $R_0=a\sqrt{\eta_{\rm w}}/(1+\sqrt{\eta_{\rm w}})$.

Neglecting the Coriolis force, the asymptotic bow shock has an opening angle $\theta_\infty$, which is related to $\eta_{\rm w}$ through $\theta_\infty-\tan\theta_\infty=\pi/(1-\eta_{\rm w})$  \cite{Canto+1996ApJ...469..729C,Wadiasingh+2017ApJ...839...80W}. If the shock surface is always optically thick to pulsar radio emission, then $\cos\theta_\infty+\sin i\cos[\pi(\phi_{\rm eg}-\phi_{\rm in})]=0$. With these two assumptions, we calculated the ratio of the two-wind ram pressures $\eta_{\rm w}$ for various orbital inclination angles $i$, finding 0.018, 0.050, and 0.24 for $i= 90^{\circ}$, $60^{\circ}$ and $30^{\circ}$, respectively. 

In reality, the Coriolis force twists the bow shocks, which leads to an asymmetric eclipse range about the conjunction phase, and the shock surface is not always optically thick. Therefore, the ratio $\eta_{\rm w}$ is probably higher than those calculated above, which was shown by equation~26 in \cite{Canto+1996ApJ...469..729C}. For the PSR~J1928+1815 binary, assuming that $\eta_{\rm w}$ is 0.25 and $\sin i=1$ and the observed eclipse orbital phase range of $\phi_{\rm eg}-\phi_{\rm in}=0.17$, we can only get such an eclipse by enlarging the companion wind region to more than the outermost optically thick shock lines from the connection line of two stars in  Fig.~\ref{geometry}.

\subsubsection*{Magnitude upper limits of the companion star}

Table~\ref{magnitude} compares the upper limits on the optical and infrared magnitudes to the expected brightness of a helium star companion at the distance of 8~kpc, given the anticipated interstellar extinction.

For the helium star companion, we theoretically estimate the visual magnitudes in different bands for three groups of stars: modeled naked helium stars without hydrogen envelop, modeled stripped helium stars with hydrogen envelop \cite{Gotberg+2018A&A...615A..78G}, and observed stripped helium stars (using the luminosity from reference \cite{Gotberg+2023ApJ...959..125G} and assuming that they have a hydrogen envelope as large as their Roche lobes). We calculated a stellar evolution model using the package
the Modules for Experiments in Stellar Astrophysics (\textsc{Mesa}, version 12115) \cite{2011ApJS..192....3P, Paxton+2019ApJS..243...10P}, from which we determined a radius - mass relation $R_{\rm He}/ R_\odot=0.205~(M_{\rm He}/M_\odot)^{0.786}$  and find the effective temperature $T_{\rm eff}=49.6~{\rm kK}(M_{\rm He}/M_\odot)^{0.495}$ for naked helium stars with the helium abundance -- the mass fraction for helium element -- $Y=0.98$ and the metallicity -- the mass fraction for all heavier elements -- of $Z=0.02$ at the age of $1~\rm million years$, where $R_{\rm He}$ is the radius and $M_{\rm He}$ is the mass of a helium star. Fitting previous simulations of modeled stripped helium stars \cite{Gotberg+2018A&A...615A..78G} we obtained a radius --  mass relation of $R_{\rm He}/R_\odot=0.331~(M_{\rm He}/M_\odot)^{0.628}$ and found that the effective temperature is related to the mass by $T_{\rm eff}=42.7~{\rm kK}(M_{\rm He}/M_\odot)^{0.458}$. Based on the observed data for stripped stars [\cite{Gotberg+2023ApJ...959..125G}, their table 3], we get the relation between the helium star luminosity ($L_{\rm He}$) and the helium star mass as  ${\rm log}(L_{\rm He}/L_{\odot})=3.45^{+1.05}_{-0.98}{\rm log}(M_{\rm He}/M_{\odot})+2.41^{+0.75}_{-0.77}$. We take the Roche-lobe radius calculated from Eq.~\ref{rRL} for $M_{\rm p}=1.4~M_\odot$ as the upper limits of the helium star radius, then determine the upper limits of visual magnitudes. Adopting the AB magnitude system \cite{fuku96}, without considering any extinction, we got the magnitude ranges for the modeled naked helium stars with a mass of $1.0$ to $1.6~M_\odot$ to be 17.2 to 16.1 mag, 17.7 to 16.6 mag, 18.1 to 17.0 mag, 18.4 to 17.4 amg, and 18.7 to 17.6 mag in the $grizy$ bands for the distance at 8.0 kpc, respectively. In the Vega magnitude system \cite{bess98}, a helium star at this distance would have a magnitude of $18.1$ to $17.0$, $18.2$ to $17.2$, and $18.3$ to $17.3$~mag in the $JHK$ bands, respectively (Table~\ref{magnitude}). For the modeled stripped stars and observed stripped stars with a hydrogen envelope, they are slightly brighter than the modeled naked helium stars. 

The binary system is located at the Galactic longitude of $l=53^\circ.29$ and the Galactic latitude of $b=0^\circ.46$. This places it close to the Perseus spiral arm at about 9.7~kpc or at the outer edge of the Sagittarius arm at about 7~kpc, just behind its tangent point where the extinction is severe. Most 3D dust maps do not extend as far as 8~kpc, except the dust model \cite{Marshall+2006A&A...453..635M} derived from the 2MASS data and does extend to a distance of about 10~kpc, which gives We the extinction in the $grizy$ bands and the $JHK$ bands as 17.7, 13.2, 9.9, 7.8, 6.4, 4.0, 2.4 and 1.5~mag (Table~\ref{magnitude}), respectively, for a star at 8~kpc. However,  the distance we derived from the DM has a large uncertainty of 30\%. In addition, circumstellar dust formed from the material ejected from the common envelope could produce additional extinction  \cite{Lu+2013ApJ...768..193L, Iaconi+2020MNRAS.497.3166I}.

Based on the non-detection of the companion we determine a tighter constraint on its mass. Helium stars follow a mass-luminosity relationship $\log (M_{\rm He}/M_\odot)\approx0.32\log(L_{\rm He}/L_\odot)-0.80$ [\cite{Gotberg+2023ApJ...959..125G}, their equation 2]. The infrared counterpart of the companion must be fainter than 18.8 mag in $K-$band minus an extinction of 1.5 mag, from which the radius upper limit of the helium star was calculated according to the black-body radiation for a given helium star mass, approximated by $\log(R/R_\odot)=-0.571\log(M/M_\odot)-0.417$ in the mass range of 1.0 to 1.6 $M_\odot$ (Fig.~\ref{mc-r}).

\subsubsection*{Formation and evolution of the compact binary}
\label{simu-popu}

After the formation of the original neutron star, the progenitor of the binary at stage I in Fig.~\ref{evolution} evolves into an intermediate/high-mass x-ray binary at stage II.  The donor star expands in response to the mass loss, leading to a dynamically unstable mass transfer \cite{Ge+2010ApJ...717..724G,Ge+2020ApJS..249....9G,Ge+2023ApJ...945....7G} and the formation of a common envelope at stage III. 

In this section, we first investigate the expected number of such compact binaries using binary population synthesis study by using modified \textsc{Bse} code
\cite{hurl02}. We then discuss the constraints of the common envelope evolution phase and model the common envelope evolution of binaries using \textsc{Mesa} \cite{2011ApJS..192....3P, Paxton+2019ApJS..243...10P}, and finally discuss the fate of such a binary system. 

\paragraph*{Binary population synthesis}

We used the modified population synthesis code \textsc{Bse}  \cite{hurl02, 
shao14, Shao+2019ApJ...886..118S} to calculate expected number of binaries similar to PSR J1928+1815 that could be observed. 
We simulated $10^7$ binary systems, assuming all binary radio pulsars with a helium star companion are formed via the evolution of isolated primordial circular binaries, beginning with zero-age main sequence stars with Sun-like metallicities ($Z = 0.02$). The physics processes include the evolution of single stars and binary star interactions such as mass transfer, accretion via stellar winds, RLO, common envelope evolution, neutron star formation, supernovae kicks, and orbital angular momentum loss due to gravitational wave radiation and magnetic braking. The initial binaries starting the evolution had these parameter distributions: the masses of the primary stars are assumed to follow a Kroupa initial mass function \cite{Kroupa+1993MNRAS.262..545K}, the mass ratios of the secondary to the primary stars follow a flat distribution between 0 and 1 \cite{Kobulnicky+2007ApJ...670..747K}, the orbital separations are uniformly distributed in logarithmic space \cite{Abt+1983ARA&A..21..343A}. The star formation in the Milky Way is assumed to have a constant rate of 3\,$M_{\odot}\,{\rm yr}^{-1} $ for 10 Gyr. 
Neutron stars are formed through core-collapse supernovae or electron-capture supernovae. The kick velocities of newly-formed neutron stars are expected to disrupt a fraction of binary systems and then influence the formation of PSR~J1928+1815-like binaries. The kick velocities are assumed to follow a Maxwellian distribution with two example velocity dispersions of $V_{\rm CC}$ = 265 and $320\,\rm km\,s^{-1}$ \cite{Hobbs+2005MNRAS.360..974H, Igoshev2020MNRAS.494.3663I} if they are formed via core-collapse supernovae, or $V_{\rm EC} = 40$ and $80\,\rm km\,s^{-1}$ if formed via electron-capture supernovae \cite{Verbunt+2017A&A...608A..57V,Igoshev2020MNRAS.494.3663I}.
We choose three example efficiencies of common envelope ejection,  $\alpha_{\rm CE}=0.3$, 1.0, and 3.0 for three sets of simulations. 

From the simulation outputs, we identified the binary systems with properties similar to PSR~J1928+1815, as detached neutron star binaries with a companion mass of $1.0\, M_\odot < M_{\rm He} < 2.0\, M_\odot$ and an orbital period of $2.4 {\rm\,hr} < P_{\rm orb} < 4.8 {\rm\, hr}$. Our calculations show that the formation rate of such a binary depends on the choices of common envelope ejection efficiencies and neutron star kick velocities. For the values we assumed above, the predicted formation rates of PSR~J1928+1815-like systems are (1.3 to 7.2) $\times 10^{-6}~\rm yr^{-1}$ (Table~\ref{birthrate and number}). In the Milky Way, we found that about 16~to~84 such binaries exist at any given time. 

If the helium star develops an expanded envelope during the helium shell burning phase and transfers matter to the neutron star via a new RLO, a proto-white dwarf may be formed. If the white dwarf has a burning residual envelope, that could cause the pulsar eclipses seen in PSR~J1928+1815. Such a post-Case BB binary is much less common because the proto-white dwarf has a very short lifetime. We calculate that there are less than 0.035 such binaries with the orbital and mass properties similar to the PSR~J1928+1815 binary in the Milky Way.

\paragraph*{Common envelope evolution and progenitors}

The  outcome of common envelope evolution depends on the progenitor mass and evolutionary stage, companion mass,  binding energy, and common envelope ejection efficiency. We investigate the common envelope ejection process from the progenitor of PSR~J1928+1815 binary by covering possible parameter spaces of progenitor mass, RGB evolutionary stage, and neutron star mass with different common envelope ejection efficiencies, then constrain the parameter spaces of the helium star progenitor. 

During common envelope evolution phase, the orbital evolution of the neutron star inside the progenitor envelope of the helium star is driven by the dynamical friction drag. With a gradual spiral-in of the neutron star, the released orbital energy $\Delta E_{\rm orb}$ must overcome the binding energy $E_{\rm bind}$ of the envelope of the donor star with an efficiency $\alpha_{\rm CE}$ to eject the envelope,  
$\alpha_\mathrm{CE} \Delta E_\mathrm{orb} = E_\mathrm{bind}$
\cite{Webbink+1984ApJ...277..355W,kool90}
. 
The orbital energy change for this system is
\begin{equation}
	\Delta E_\mathrm{orb} =   \frac{G M_\mathrm{i} M_\mathrm{p}}{2 a_{\mathrm{i}}}  - \frac{G M_\mathrm{f} M_\mathrm{p}}{2 a_{\mathrm{f}}},
	\label{eorb}
\end{equation}
where $G$ is the gravitational constant, $M_\mathrm{i}$ and $M_\mathrm{f}$ are the initial and final mass of the donor, $a_\mathrm{i}$ and $a_\mathrm{f}$ are the initial and final semi-major axes, such that $a_\mathrm{f} \ll a_\mathrm{i}$. Following previous work \cite{Ge+2010ApJ...717..724G,Ge+2022ApJ...933..137G,Ge+2024ApJ...961..202G}, we express the binding-energy with more considerations as
\begin{equation}
		E{'}_{\mathrm{bind}} =\int_{0}^{M_{\mathrm{i}}}\left(-\frac{G m}{r}+U\right) {\rm d} m - \int_{0}^{M_{\mathrm{f}}}\left(-\frac{G m}{r}+U\right) {\rm d} m,
\label{Ebind2}
\end{equation}
where $m$ is the varying mass and $r$ is the radius of the donor star, and $U$ is its specific internal energy including the plasma recombination and molecules formation from atoms \cite{Ivanova+2013A&ARv..21...59I}.  This approach includes progenitors with non-degenerate cores, and address the remnant response to the common envelope ejection and the effect of different companion masses \cite{Ricker+2012ApJ...746...74R,Andrews+2015ApJ...801...32A}.
The final mass $M_{\mathrm{f}}$ is not just the core mass, but is determined when the remnant shrinks back within its Roche lobe.
We use the common envelope efficiency parameter $\beta_\mathrm{CE}$ \cite{Ge+2022ApJ...933..137G,Ge+2024ApJ...961..202G} to relate the new binding-energy and the released orbital energy via
\begin{equation}
	\beta_\mathrm{CE} \Delta E_\mathrm{orb} = E{'}_\mathrm{bind}.
	\label{beta1}
\end{equation}

Given the final mass of the helium star companion in the range of $1.0\, M_\odot < M_{\rm c} < 2.0\, M_\odot$, its progenitor must be an RGB star with a mass of 4 to 10 $M_\odot$. The common envelope evolution with such a RGB star and a neutron star has been studied \cite{Webbink+1984ApJ...277..355W, Iben+1984ApJS...54..335I, Livio+1988ApJ...329..764L}. 
To reproduce the evolution of this compact binary from stage III to stage IV, we perform simulations for the common envelope ejection process using an adiabatical mass-loss model \cite{Ge+2022ApJ...933..137G, Ge+2024ApJ...961..202G}. The RGB progenitors have a mass of $4~\rm to~10\,$ $M_\odot$ and  metallicity of $Z = 0.02$, and neutron stars have a mass of 1.0 to 2.5~$M_\odot$, evolving in common envelope with an efficiency parameter of $\beta_\mathrm{CE} = 1.0$ and 0.5  (Fig.~\ref{ge-cee}).  

We find that progenitor systems merge if the progenitor is in the early stage of RGB. Survived systems have different orbital periods and companion masses (Fig.~\ref{ge-cee}), depending on $\beta_\mathrm{CE}$, neutron star mass $M_{\rm NS}$, and the mass $M_{\rm i}$ and radius of a RGB star when common envelope phase initiates.
For the cases with $\beta_\mathrm{CE}=1.0$, with a neutron star of 1.44\,$M_\odot$, the common envelope phase initiates when an RGB star has a mass of 6.3\,$M_\odot$ and a radius of 130.6\,${R}_\odot$; it ends with a 1.23\,$M_\odot$ helium star and an orbital period of 7.8~hours.
Evolution of such an RGB star with a neutron star of 1.30\,$M_\odot$ in a binary ends with a helium star, and the mass is 1.11\,$M_\odot$ and the final orbital period is 2.9 hours.  
For a smaller $\beta_\mathrm{CE}=0.5$ (Fig.~\ref{ge-cee}B), common envelope phase initiates when an RGB star has a mass of 6.3\,$M_\odot$ and a radius of 163.0\,${R}_\odot$ for the neutron star of 1.8\,$M_\odot$, and ends with a 1.20\,$M_\odot$ helium star and an orbital period of 6.3 hours. For a neutron star of 1.6\,$M_\odot$, the helium star mass is 1.14\,$M_\odot$ and the final orbit period 
is 1.1 hours.

We find from two sets of simulations  (Fig.~\ref{ge-cee}) that for the binary system of PSR~J1928+1815 with a 3.6-hour orbital period and the companion of 1 to 2\,$M_\odot$-helium star, the most likely RGB progenitor has a mass of 6.3\,$M_\odot$ (Fig.~\ref{Pb-m2}) and the final companion star of PSR~J1928+1815 has a mass $\lesssim$1.2 $M_\odot$. 
If the progenitor RGB star has a mass heavier than $8\, $$M_\odot$, the final orbital period after the common envelope ejection is much longer; 
if the RGB star has a mass less than $5\, $$M_\odot$, the helium star mass is less than 1~$M_\odot$, lower than the companion mass of PSR J1928+1815.
Given the numerous uncertainties inherent in common envelope evolution simulations, these parameter ranges should be taken as indicative estimates rather than strict constraints.

\paragraph*{Modeling the common envelope evolution} \label{Method_CEsample}

We use the stellar evolution code \textsc{Mesa} \cite{2011ApJS..192....3P, Paxton+2019ApJS..243...10P} to simulate the evolution of high-mass x-ray binaries for the formation of PSR~J1928+1815-like binary systems. We use version 12115 of \textsc{Mesa} which simulates the common envelope evolution by implementing a method to determine the core–envelope boundary self-consistently \cite{2021A&A...650A.107M}. 

The binary systems initially contain a neutron star of $1.4~M_\odot$ and a companion star of 8 to 20~$M_\odot$, with initial orbital periods in a range of  300 to 3000 days. The neutron star is treated as a point mass, and the companion donor star is initially set to be a zero-age main-sequence star with solar metallicity $Z =0.02$. The hydrogen and helium elements of the donor star have mass fractions of 0.70 and 0.28, respectively. When the mass-transfer rate $\dot{M}_{\rm tr}$ from the donor star to the neutron star during RLO reaches a threshold of $\dot{M}_{\rm CE} \simeq 1.0\,M_{\odot}\,\rm yr^{-1}$, the binary is supposed to evolve into a common envelope phase \cite{2021A&A...650A.107M}. 
The binary system evolves under an energy prescription \cite{Webbink+1984ApJ...277..355W}, and the efficiency of common envelope ejection is set in the range of $\alpha_{\rm CE}=0.1$ to 3.0.

During a common envelope phase, the mass-transfer rate is related to the donor's radius $R_{\rm c}$ and its $R_{\rm L}$ \cite{2021A&A...650A.107M}. If $R_{\rm c} > R_{\rm L}$, we assume a high mass-loss rate  $\dot{M}_{\rm high} = \dot{M}_{\rm CE}$, which approximately corresponds to the mass loss on the adiabatic timescale of the envelope. If $0.98 R_{\rm L} < R_{\rm c} < R_{\rm L}$, we interpolate a mass transfer 
rate between $\dot{M}_{\rm high}$ and a low mass transfer rate $\dot{M}_{\rm low} = 10^{-5}\,M_{\odot}\,\rm yr^{-1}$, 
here $\dot{M}_{\rm low}$ is close to that for the mass transfer occurring on the nuclear timescale of the donor star. The common envelope phase ends when $R_{\rm c} <  
0.98 R_{\rm L}$. 

The neutron star mass $M_{\rm NS}$ is assumed to increase by a negligible amount during common envelope evolution, and mass accretion onto the neutron star is assumed to be limited by the Eddington rate of $\dot{M}_{\rm Edd}=2 \times 10^{-8}M_{\odot}\,\rm yr^{-1}$.  Each simulation is terminated if either the core carbon of the donor star is depleted or a merger between the donor star and the neutron star happens. 

Fig.~\ref{CE-EwT} (A-D) show example simulation results. In the case of $\alpha_{\rm CE}=1.0$, all post-common envelope binaries, if they survive, have orbital periods of $>1$~day. In the case $\alpha_{\rm CE}=0.3$,  some close binaries similar to PSR~J1928+1815 are formed (Fig.~\ref{CE-EwT}A), for example, a binary initially consisting of a $1.4M_\odot$ neutron star and an $8M_\odot$ donor star in a $\sim720$-day orbit. Up to $42.58\,\rm million years$, the donor star has lost $\sim 0.7\,M_{\odot}$ from the envelope via stellar winds, and the neutron star accreted about $\sim 0.005\,M_{\odot}$ from the companion. 

After $42.58\,\rm million years$, the donor star evolves onto the asymptotic giant branch (AGB), and the RLO mass transfer starts. The mass transfer rapidly increases to $1.0\,M_{\odot}\,\rm yr^{-1}$ and then the common envelope phase starts. The binary orbital period quickly decreases to $\sim 4.5$ hours. During the common envelope phase, a mass of about $\sim 4.8\,M_{\odot}$ has been stripped from the envelope by the neutron star, but a mass of $\sim 0.04\,M_{\odot}$ of the hydrogen envelope remains when the donor star radius falls to $R_{\rm c} < 0.98
R_{\rm L}$ at the end of common envelope evolution. 
Afterward, another stable mass transfer phase starts and lasts for $\sim 0.05$ million years, with a varying mass transfer rate of $10^{-6}$ to $10^{-4}M_{\odot}\,\rm yr^{-1}$, corresponding to an ultraluminous x-ray source \cite{2019ApJ...883L..45F}. At the end of this phase, the mass-transfer rate drops below $10^{-10}\,M_{\odot}\,\rm yr^{-1}$ due to the contraction of the donor star, and about $\sim 0.002\,M_{\odot}$ material is accreted by the neutron star. At this stage, the system could be shown as a PSR~J1928+1815-like binary. The neutron star mass continues to grow due to wind accretion, and the pulsar binary system can live for a few 10$^4$ yr.  At the time of $42.64$ million years, the donor star expands again to fill its Roche lobe, the binary undergoes a very short mass transfer phase and then the simulations are terminated due to core carbon depletion. 

We conclude that the common envelope evolution with an AGB star is capable of producing systems similar to the PSR~J1928+1815 binary with a surviving time of only $10^4$ yr.

\paragraph*{Other possibilities for PSR~J1928+1815 companion}
\label{subsec2-7}

Our observations and theoretical constraints from stellar evolution theory indicate that the PSR~J1928+1815 binary consists of a pulsar with a helium star companion after a common envelope stage (Fig.~\ref{evolution}). The helium star most likely has a mass between 1.0 to 1.6 $M_\odot$ (Fig.~\ref{mc-r}), but a higher helium star mass cannot be completely excluded mainly because of the uncertainties of pulsar distance, dust extinctions, and theoretical parameters for the common envelope evolution phases. If the helium star companion has a mass greater than 2.8~$M_\odot$, the companion could experience an ultra-stripped supernova to collapse to a neutron star  \cite{jiang2021,Guo+2024MNRAS.530.4461G}.

As discussed above, the companion of PSR~J1928+1815 could be a massive proto-white dwarf with a non-degenerate residual envelope. A proto-white dwarf companion has been found in the eclipsing redback pulsar PSR~J1816+4510 \cite{Kaplan+2013ApJ...765..158K}, but its companion mass is only 0.19 $M_\odot$. The lifetime of a proto-white dwarf, $\Delta t_{\rm proto}$, depends on its mass, $M_{\rm WD}$, decreasing rapidly following $\Delta t_{\rm proto}\sim400~{\rm million years}~(0.2 M_\odot/M_{\rm WD})^{7}$ for a proto-helium white dwarf \cite{Istrate+2014A&A...571L...3I}. 
The minimum companion mass of our PSR~J1928+1815 is 1.0 $M_\odot$, and the lifetime of such a proto-white dwarf should be similar to the time scale for the Roche lobe decoupling, $\lesssim$10$^4$~yr \cite{Tauris+2012MNRAS.425.1601T}. Because of the short lifetime, the probability of observing such a system is small (Table~\ref{birthrate and number}). 

To investigate the detection probability, we simulate the evolution of a neutron star binary with a helium star of 2.0~$M_\odot$ in an orbit with a period of 0.1~day (Fig.~\ref{CE-EwT} E-H) by using \textsc{Mesa} \cite{2011ApJS..192....3P,Paxton+2019ApJS..243...10P} with same physical assumptions as the above section. After the mass-transfer phase of the Case BB RLO, this system evolves into a pulsar binary with an orbital period of $\sim$0.16 days, in which the companion star is a proto-white dwarf consisting of a 1.03 $M_\odot$ carbon-oxygen core and a 0.016 $M_\odot$ helium envelope. The helium shell burns stably on the surface, instead of causing a shell flash, due to the high core temperature. As the helium in the companion surface is burned into carbon and oxygen, the radius of the proto-carbon-oxygen white dwarf gradually decreases to 0.02~$R_{\rm L}$, which cannot produce the eclipses in a large orbit phase range we observed. The time scale for such a process is  $\lesssim$3$\times$10$^4$ yr (Fig.~\ref{CE-EwT}G). It is therefore unlikely to detect such a short-lived proto-white dwarf with large remnants. 

When the initial helium star mass changes to 2.4~$M_\odot$, at the end of Case BB RLO, the detached companion star evolves to a helium star with a thick helium envelope of 0.12 $M_\odot$ and a carbon-oxygen core of 1.15 $M_\odot$ that is massive enough to ignite the burning of carbon-oxygen. Such an evolved helium star with a carbon-oxygen core could produce the observed eclipses. However, the carbon-oxygen core burns so rapidly that such a binary system lasts $\lesssim2\times10^4$~yr before the companion expands again (Fig.~\ref{CE-EwT} I-L), initiating another mass transfer phase in Case BBB RLO during the carbon-shell burning \cite{Tauris+2023pbse.book.....T}. %
Afterwards, the companion becomes a short-lived proto-white dwarf. Compared to the helium main sequence stars with a lifetime of millions of years, such a binary system with an evolved helium star companion is unlikely to be detected. 
Analogously, a similar evolved helium star could be formed via the Case C channel of binary evolution \cite{Tauris+2023pbse.book.....T} with a similar lifetime  (Fig.~\ref{CE-EwT} A-D), much shorter than that of a helium star in the core helium-burning phase.

\paragraph*{The fate of the PSR~J1928+1815 binary system}
\label{fate}

The fate of the PSR~J1928+1815 binary system depends on the mass of the companion helium star. As discussed above, the eclipses of PSR~J1928+1815 suggest the helium star has a mass of less than 2 $M_\odot$. If so, it will evolve and form a carbon-oxygen or oxygen-neon-magnesium white dwarf finally (Fig.~\ref{Pb-m2}). Using the \textsc{Mesa} \cite{Paxton+2019ApJS..243...10P}, we performed additional binary evolution simulations for such a neutron star -- helium star system to predicted the fate of the {binary system. 

For the companion, a typical Population I metallicity ($\rm Z = 0.02$) is adopted, in which the initial helium star is composed of helium with $Y = 0.98$ and Sun-like metallicities ($Z$ = 0.02) \cite{Wang+2021MNRAS.506.4654W}. The neutron star is set as a point mass, and the fraction of transferred material accreted by the neutron star is set to be 30\%. The initial binary model starts with a helium star with a mass of 1.15~$M_{\odot}$ and a neutron star with a mass of 1.4~$M_{\odot}$ in orbit with a period of 3.6~hr, exactly as as observed for the PSR~J1928+1815 system. The predicted binary evolution is shown in Fig.~\ref{casebb_yl}. We predicted the initial helium star will undergo helium-core burning lasting for $\sim11.6\, \rm million years$. After the exhaustion of the helium in the core, the envelope of the helium star expands. After about $0.56\,\rm million years$, the helium star begins to fill its Roche-lobe, entering a new RLO phase. The binary mass-transfer rate, again, exceeds the Eddington rate of helium accretion ($\sim4\times10^{-8}~M_\odot$\,yr$^{-1}$) and lasts for $\sim 0.40\rm \, million years$. In this phase, we predict the binary system could appear as an ultraluminous x-ray source \cite{Shao+2019ApJ...886..118S}. Due to the rapid mass transfer, the helium star of 1.15~$M_\odot$ eventually evolves to a carbon-oxygen white dwarf with a mass of 0.82~$M_\odot$ (Fig.~\ref{Pb-m2}). The binary orbit period first decreases to $\rm3.53\, hr$ because of gravitational wave radiation, and then increases to $\rm6.21\,hr$ due to the binary mass transfer. After the mass-transfer phase (stage V in Fig.~\ref{evolution}), the neutron star can accrete material of $\sim 0.013\, M_\odot$, and is spun up to $\rm 5.67\,ms$ (according to Eq.~\ref{Meq}).

\subsubsection*{Pulsar recycling }\label{sect_recy}

The neutron star could have been spun up before, during, or after the common envelope phase. The lifetime of these phases and the amount of accreted mass are very different. PSR~J1928+1815 has a period of 10.55~ms, recycled by the accreting mass from the companion. The amount of accreted mass ($M_{\rm acc}$) needed to spin up a pulsar is \cite{Tauris+2012MNRAS.425.1601T}: 
\begin{equation}
    M_{\rm acc}/ M_{\odot} =0.22~\times 
     \frac{ (M_{\rm p}/M_\odot)^{1/3} } {(P/{\rm ms})^{4/3} } .
	\label{Meq}
\end{equation}
Assuming the pulsar mass $M_p$ is 1.4 $M_\odot$, at least a mass of $M_{\rm acc} = 0.01 M_\odot$ has been accreted to spin up the pulsar to $P=10.55$~ms. 

PSR~J1928+1815 has accreted some material before the common envelope evolution through wind accretion and RLO, but both processes cannot spin up a neutron star to the short spin period $P$ of PSR~J1928+1815. The wind accretion in high-mass x-ray binaries can accrete a mass of a few 10$^{-3}$ $M_\odot$,  can both spin up and spin down during accretion depending on the direction of the angular momentum carried by the wind entering the neutron star magnetosphere \cite{Tauris+2023pbse.book.....T}.  High-mass x-ray binary pulsars have such wind accretion and have periods in the range of 1 to $10^4$~s, most $>$100~s\cite{Chaty+2022abn..book.....C, Tauris+2023pbse.book.....T}, which is much longer than the spin period of PSR~J1928+1815. In the RLO phase, if the neutron star accretes at a rate no greater than the Eddington limit of a few 10$^{-8}$ $M_\odot$\,yr$^{-1}$, it requires about 0.5 million years to spin up the neutron star to 10.55 ms. RLO before the common envelope phase cannot last that long. 

The common envelope phase ends with a remnant thermal readjustment phase. It has been proposed that a neutron star can be recycled by hypercritical accretion in this phase \cite{Ivanova+2011ApJ...730...76I}. The mass transfer continues until the core shrinks faster than it expands due to mass loss, and the duration found in simulations via this Case C channel is only about 0.05 million years (Fig.~\ref{CE-EwT}B), and the mass transfer rate is 10$^{-4}~M_\odot\,\rm yr^{-1}$. 
In the post-common envelope stage, wind accretion from the helium star can provide an accretion mass of at most $\rm\sim10^{-4}~M_\odot$ \cite{Tauris+2012MNRAS.425.1601T}, insufficient to spin up the neutron star. 

If the binary has a compact orbit of $\rm P_{orb}\lesssim$2~hr, the new RLO phase may be initiated while the helium star is on its helium main sequence. This evolution scenario is Case BA RLO. Otherwise, an RLO phase will start when the  helium star expands and overfills the Roche lobe, which is Case BB RLO if the helium shell is burning, or Case BC RLO during or after the carbon core burning phase \cite{Dewi+2002MNRAS.331.1027D,Tauris+2012MNRAS.425.1601T}. Both Roche-lobe overflow phases have time scales much longer than the common envelope phase, which could be enough to spin up the neutron star  
\cite{Wang+2009MNRAS.395..847W, Tauris+2011MNRAS.416.2130T, Tauris+2012MNRAS.425.1601T, Guo+2023MNRAS.526..932G}. This mechanism has been invoked to explain compact binary systems of a pulsar and massive white dwarf including  the PSR J1952+2630 binary\cite{Lazarus+2014MNRAS.437.1485L}. 
However, the orbital period of PSR~J1928+1815 is 3.6 hr, too long to experience a Case BA RLO \cite{Dewi+2002MNRAS.331.1027D, Tauris+2012MNRAS.425.1601T}, 
and the possibility for the system to experience the Case BB RLO is very small (see Table~\ref{birthrate and number}).

For accretion in the typical duration of the common envelope phase, less than $10^3$~yr \cite{Podsiadlowski+2001ASPC..229..239P,Passy+2012ApJ...744...52P,Tauris+2012MNRAS.425.1601T}, if the accretion rate is constrained by the Eddington limit (a few $10^{-8}$ $\rm M_\odot~yr^{-1}$), the accreted mass would be negligible. In this case, evening including the accretion before the common envelope evolution phase, the pulsar cannot be spin-up much and the period is hardly less than a few tens of milliseconds (stage IVb in Fig.3).
Without invoking another new mass-transfer phase, pulsar recycling could occur only if the accretion rate was much higher than the Eddington limit for the neutron star. Such a highly super-Eddington accretion process could effectively spin up the neutron star during the common envelope phase \cite{Houck+1991ApJ...376..234H} or post-common-envelop remnant thermal readjustment phase \cite{Ivanova+2011ApJ...730...76I}, but a large amount of released gravitational potential energy has to be carried away by neutrinos that escape much easier than photons. In a dense environment like the common envelope, neutrino loss may provides the necessary cooling mechanism \cite{Houck+1991ApJ...376..234H}, and the mass accretion rate  must reach $\gtrsim10^4\dot{M}_{\rm Edd}$ \cite{MacLeod+2015ApJ...798L..19M}.   

We conclude that the PSR~J1928+1815 binary system experienced recycling of the neutron star possibly via neutrino-cooled accretion during the common envelope phase \cite{Houck+1991ApJ...376..234H}. 

\newpage


\begin{figure}
    \centering
    \includegraphics[width=0.8\columnwidth]{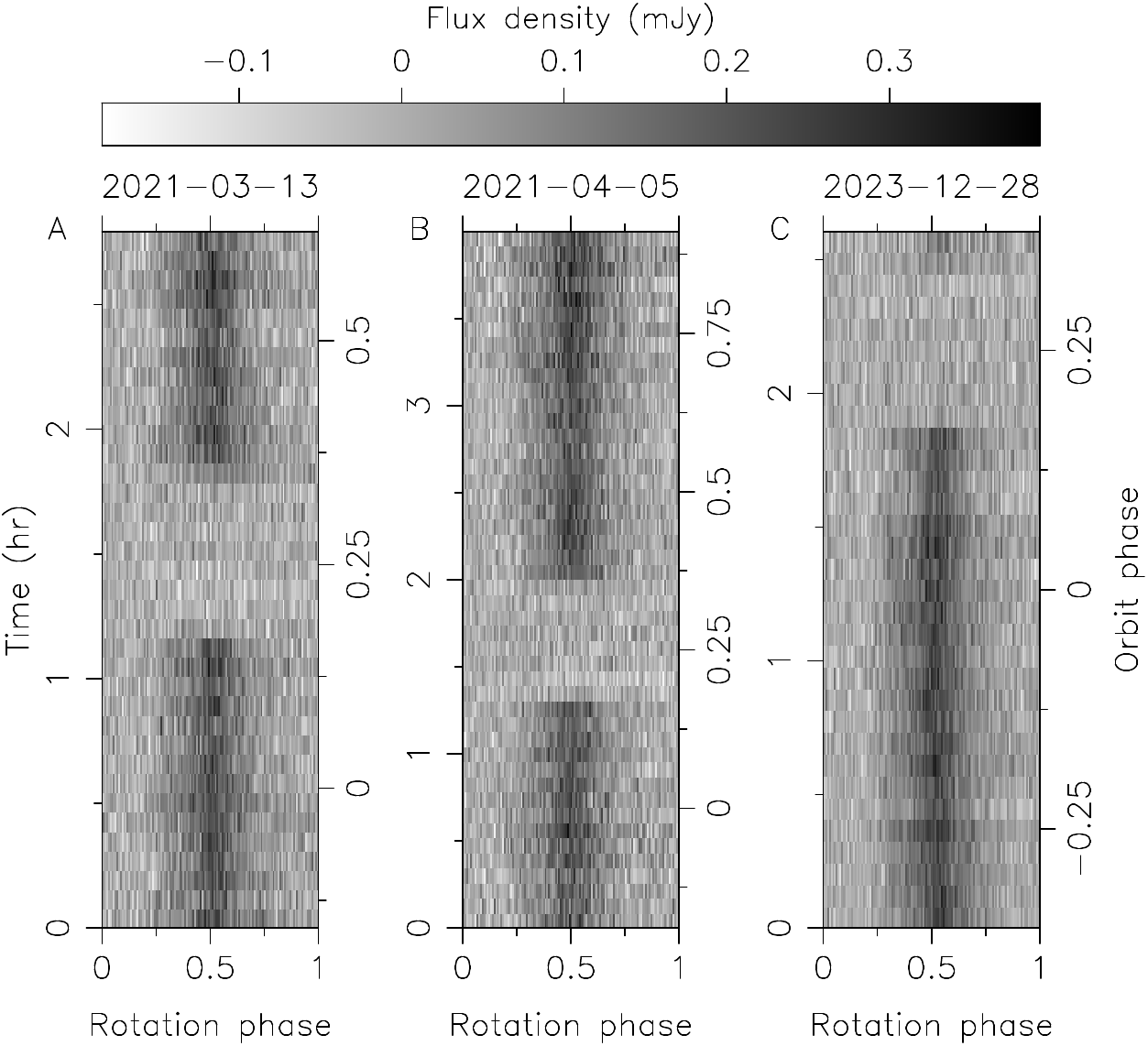}
    \caption{{\bf 2-dimensional  plots of pulsar radio emission.} Data (grey scale) are shown on the observation time -- rotation phase plane for the three long FAST observations carried out on ({\bf A}) 2021 March 13, ({\bf B}) 2021 April 5, and ({\bf C}) 2023 December 18. Eclipses appear in the orbit-phase range near conjunction $\phi_0=0.25$ with a duration of more than a half hour.
    }
    \label{time-phase}
\end{figure}

\begin{figure}
    \centering
    \includegraphics[width=0.7\columnwidth]{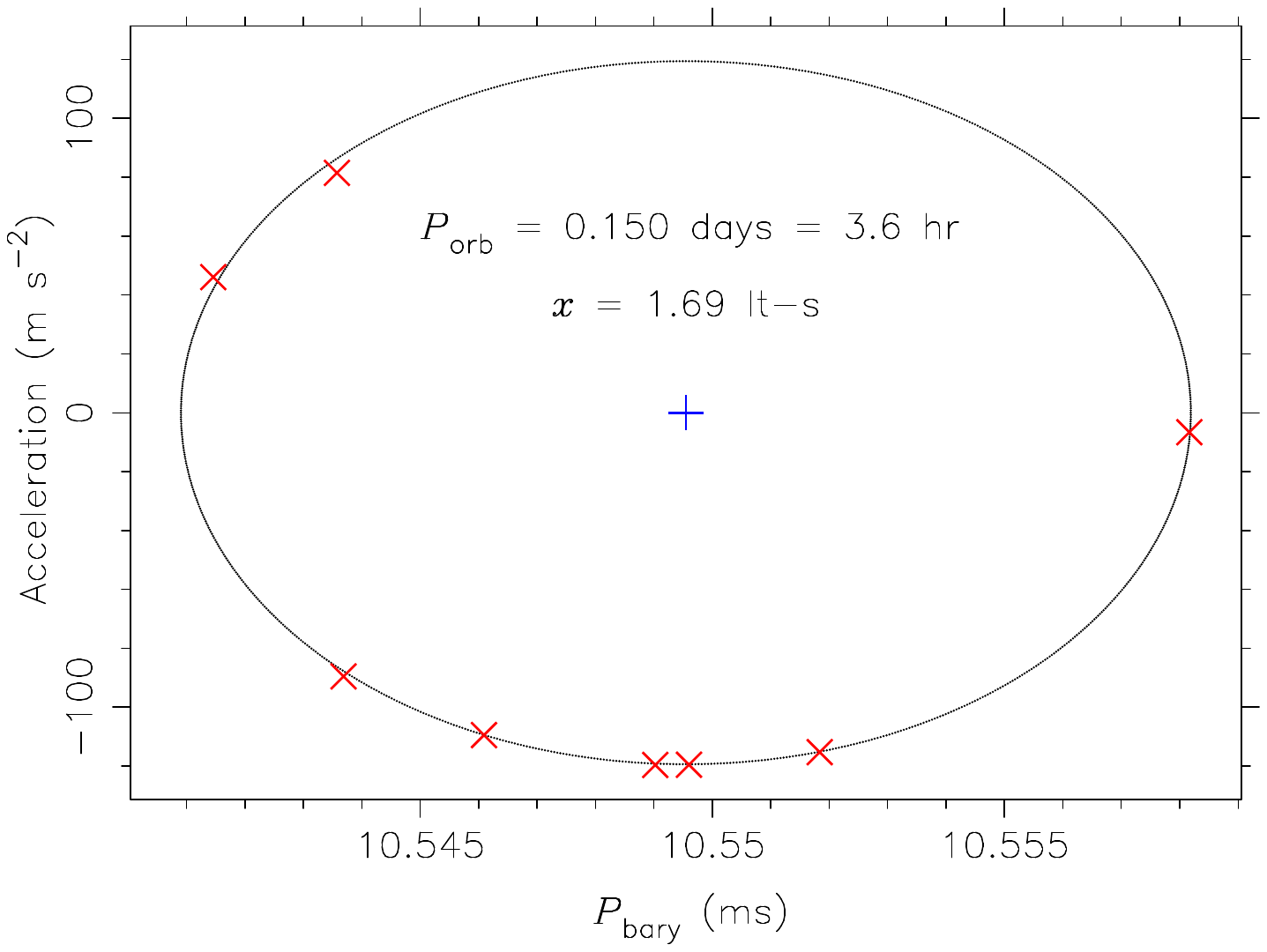}
    \caption{{\bf Determination of primary orbital parameters.} 
    We plotted crosses for the barycentric spin periods $P_{\rm bary}$ and accelerations of PSR J1928+1815 obtained from FAST observations in 2020. Following \cite{Freire+2001MNRAS.322..885F} we fitted an ellipse to the data and found the preliminary orbital period $P_{\rm orb}=0.150~{\rm days} = 3.6~{\rm hr}$ and the projected semi-major axis $x = 1.69$ lt-s. 
    }   
    \label{P0-a}
\end{figure}

\begin{figure}
\centering
   \includegraphics[width=0.49\columnwidth]{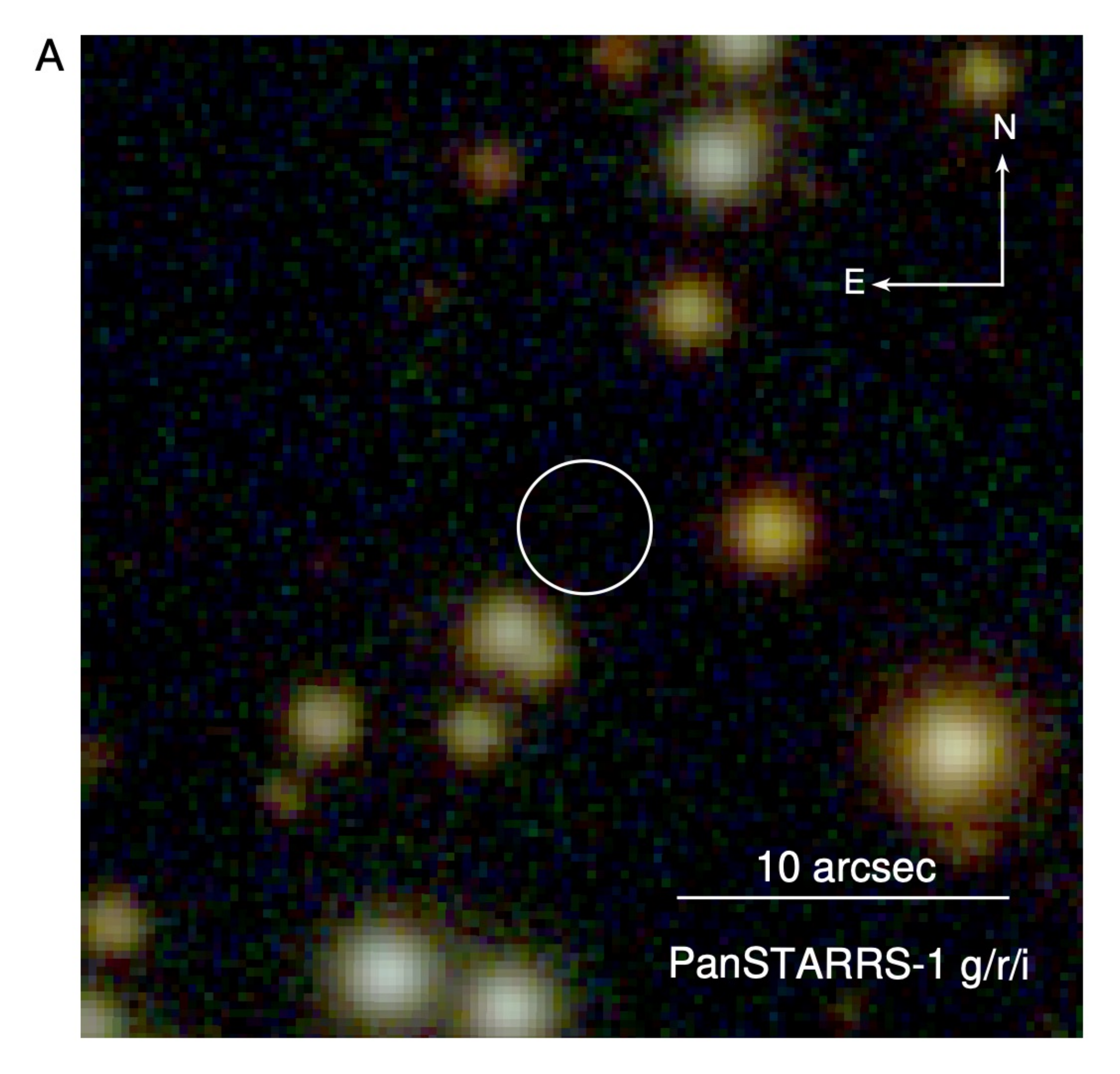}
   \includegraphics[width=0.49\columnwidth]{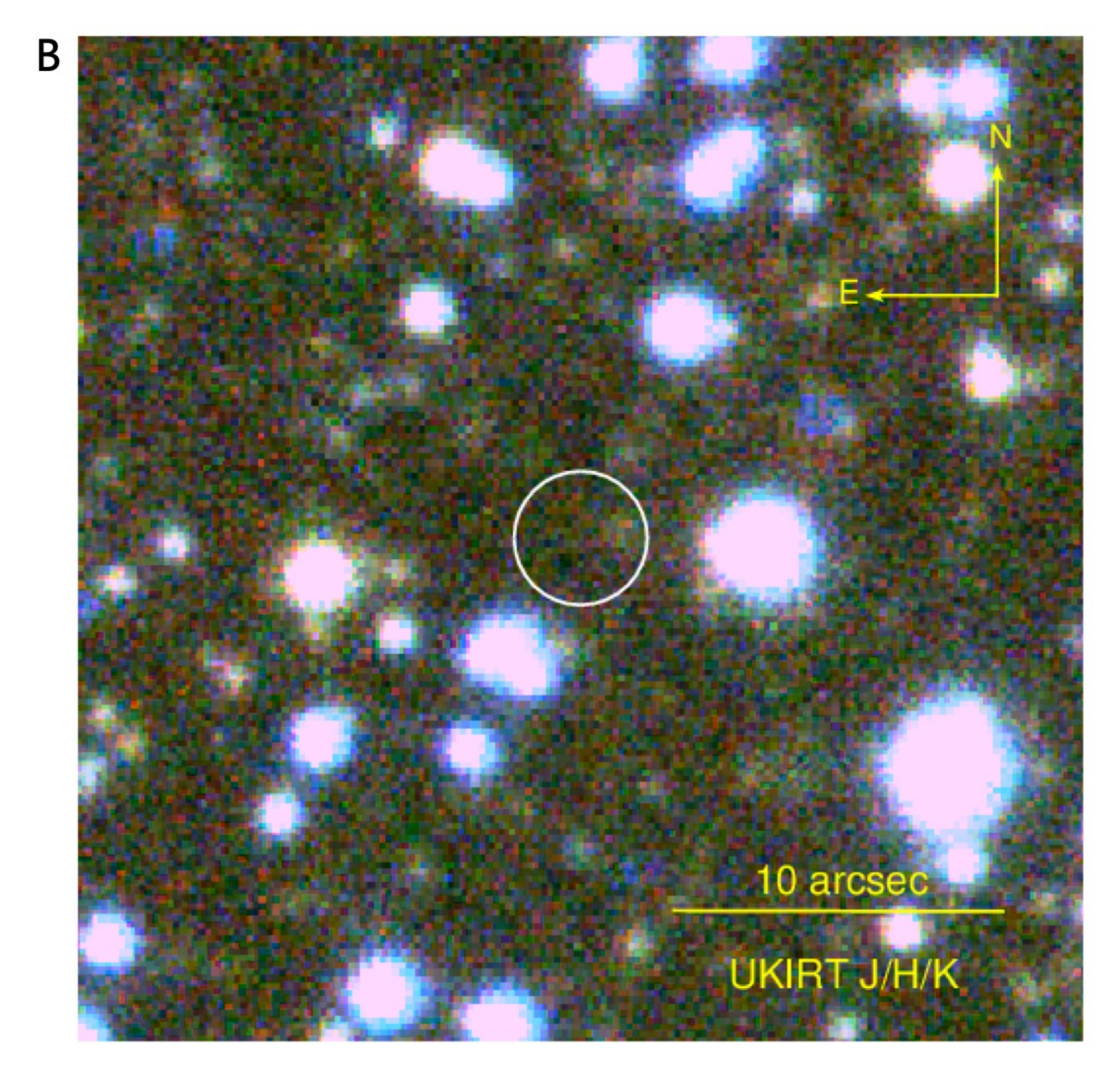}
   \caption{{\bf Optical and infrared images around PSR~J1928+1815}. ({\bf A}) a false-color Pan-STARRS1 optical image composited using the $g$ (blue), $r$ (green), and $i$ (red) bands. The white circle in both panels marks a circle with a radius of 2 arcseconds around the position of PSR~J1928+1815, within which no optical counterpart is detected. ({\bf B}) a false-color UKIRT image composited using the $J$ (blue), $H$ (green) and $K$ (red) band data from UKIDSS. No infrared counterpart is detected. The resulting magnitude limits are listed in Table~\ref{magnitude}.
   }
   \label{pans-ukirt}
\end{figure}

\begin{figure}
\centering
   \includegraphics[width=0.95\columnwidth]{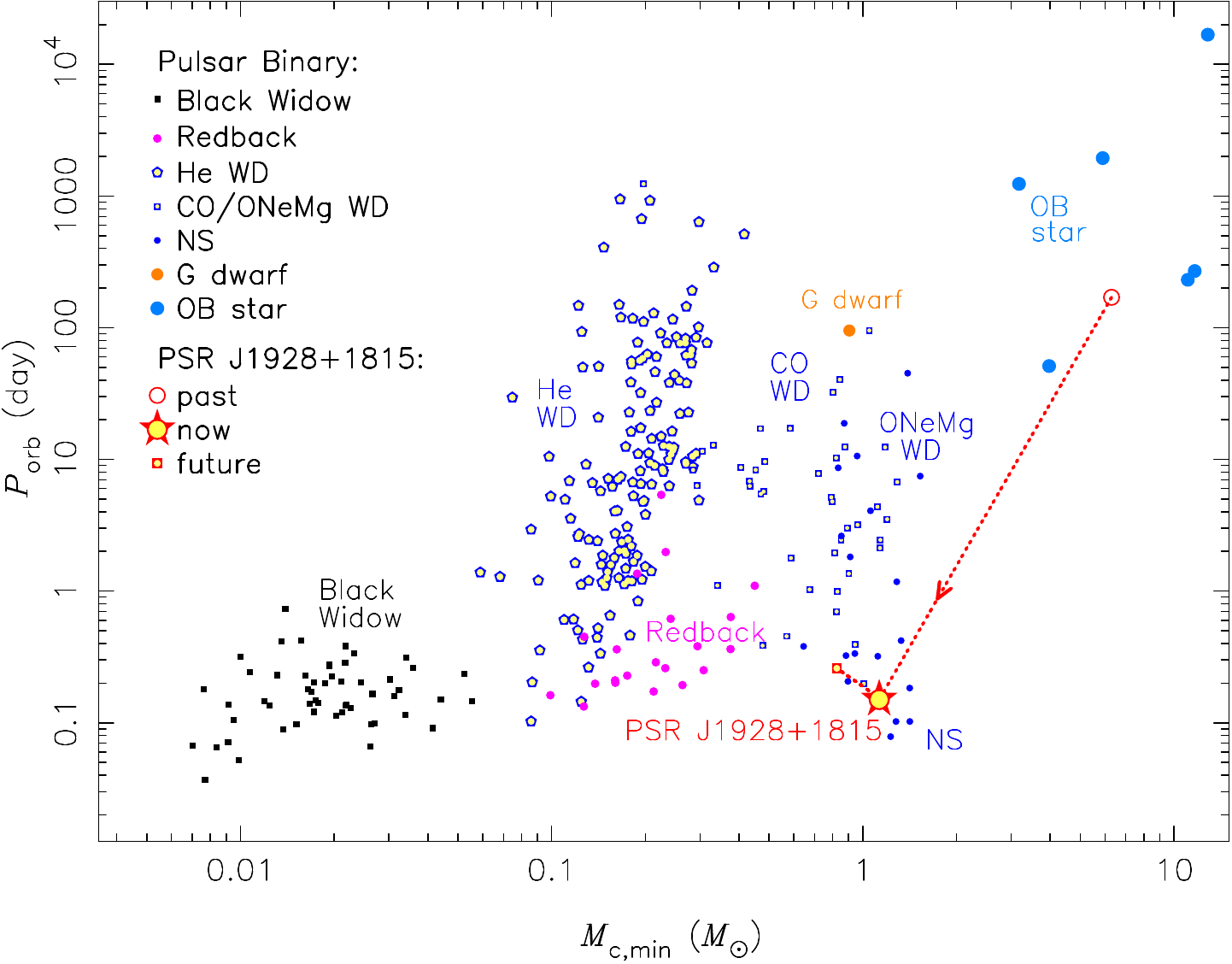}
   \caption{{\bf PSR~J1928+1815 compared to various pulsar binaries.} All pulsar data of orbital periods $P_{\rm orb}$ and minimum companion masses $M_{\rm c,min}$ for other pulsar binaries are from the Australia Telescope National Facility (ATNF) pulsar catalog version 2.6.0 \cite{Manchester+2005AJ....129.1993M}, except for PSR~J1928+1815 from this paper. Labels and symbols (see legend) indicate the nature of the companion in each system. The companion of PSR~J1928+1815 is several times more massive than those in other spider pulsars. The dotted red line indicates the orbital period changes in the post and the future suggested by our simulations (see text). 
   } 
   \label{Pb-m2}
\end{figure}

\begin{figure}
\centering
    \includegraphics[width=0.9\columnwidth]{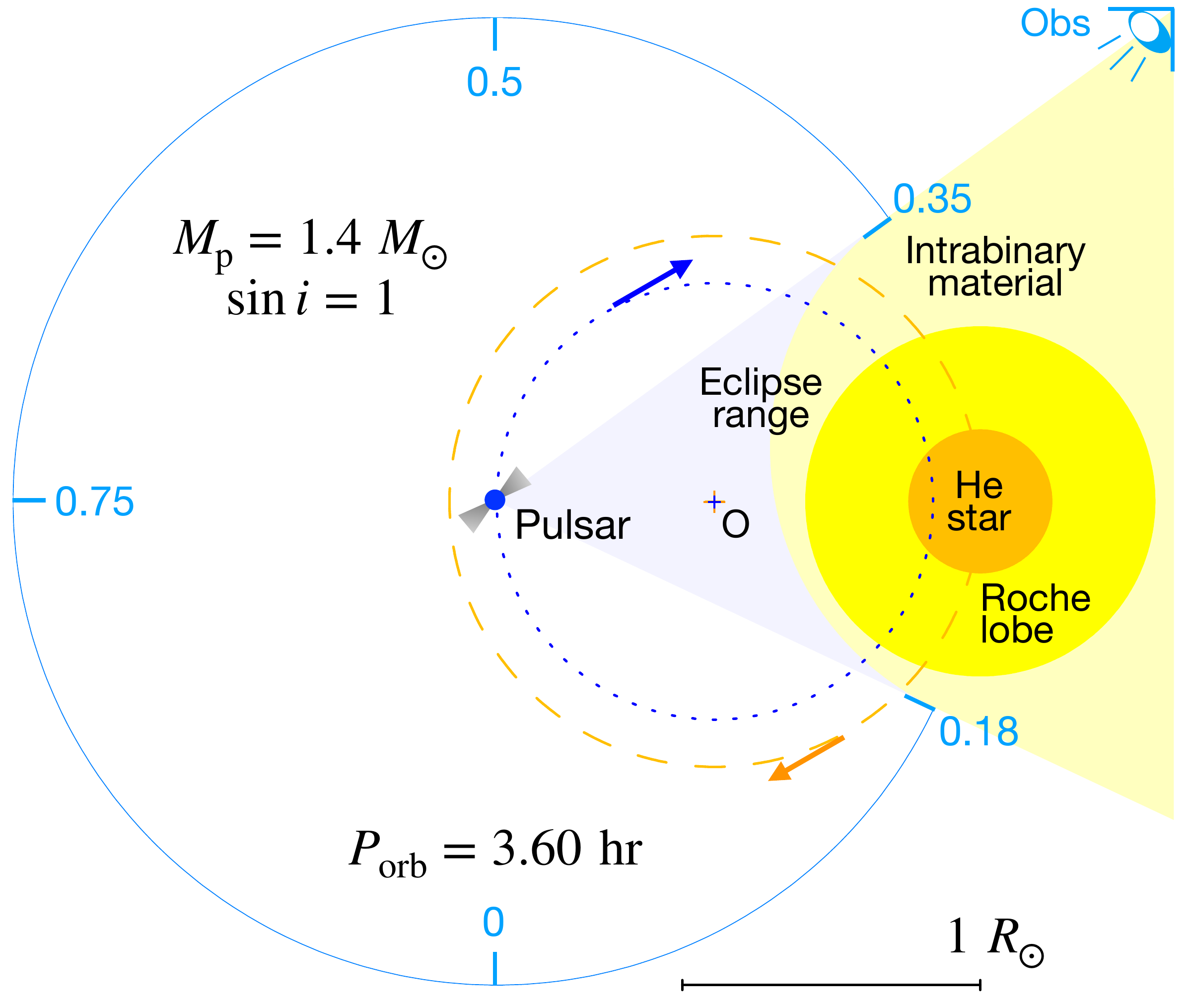}
    \caption{{\bf Illustration of the orbital geometry  we infer for the PSR~J1928+1815 binary.} The pulsar and the companion helium star  orbit each other around the barycenter (cross labeled O) on the dotted blue and dashed orange paths, respectively, which are viewed from directly above. The pulsar winds overwhelm the winds from the companion star (light yellow shading, larger than its Roche lobe region in the dark-yellow region) to form a bow shock, which causes the eclipses indicated by the light blue area. In a co-rotating frame, the pulsar and the companion are motionless in space, so the line of sight from the observer (labeled Obs) to the pulsar rotates uniformly anticlockwise with a period of $P_{\rm orb}=3.60~\rm hr$, as indicated by the outer light-blue circle.  
 }
    \label{geometry}
\end{figure}

\begin{figure}[htb!]
	\centering
	\includegraphics[width=0.98\textwidth]{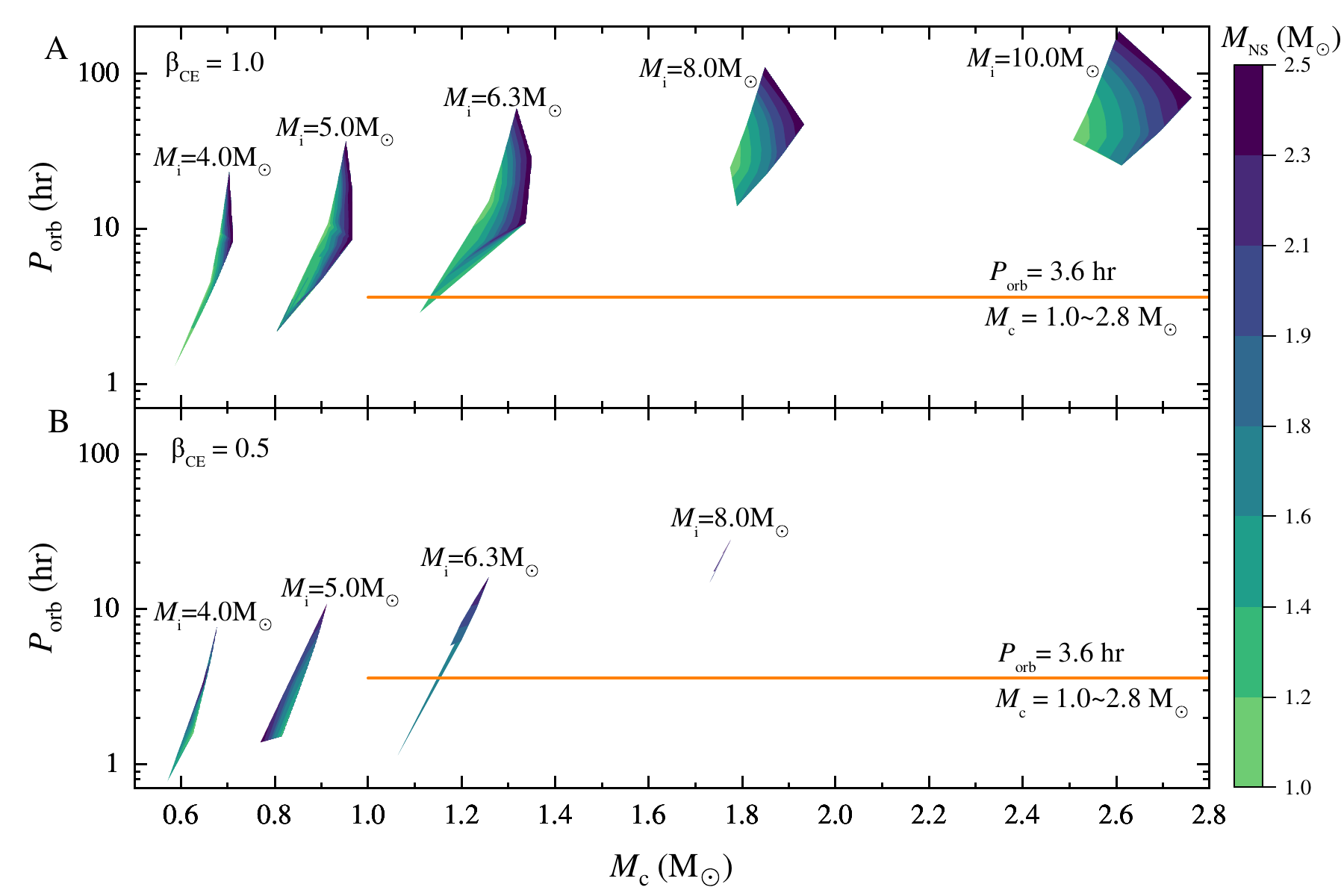}
	\caption{{\bf 
  Simulated progenitors that could produce the PSR~J1928+1815 binary via the common envelope ejection.}
 Results are shown for simulations with an common envelope ejection efficiencies of ({\bf A) }$\beta_\mathrm{CE} = 1$, and ({\bf B) $\beta_\mathrm{CE} = $}0.5, together with different initial neutron star masses $M_{\rm NS} =$ 1.0 to  2.5\,$M_\odot$ as expressed in color,  different initial progenitor RGB star masses of $M_{\rm i} = 4.0$, 5.0, 6.3, 8.0, and 10.0\,$M_\odot$. For the same set of $M_{\rm i}$ and $M_{\rm NS}$, RGB stars in different initial orbits can have different sizes when the RLO initiates, therefore binaries can evolve to a pulsar-helium star binary with different orbital periods and helium star masses indicated by contours. Only simulations with initial companion mass $M_{\rm i}$ around 6.3 $M_\odot$ can produce a binary system of PSR~J1928+1815 with an orbital period of 3.6~hr and the helium star mass of $>1.0\,M_\odot$ indicated by the orange line.
 }
 \label{ge-cee}
\end{figure}

\begin{figure}
\centering
  \includegraphics[width=0.33\textwidth]{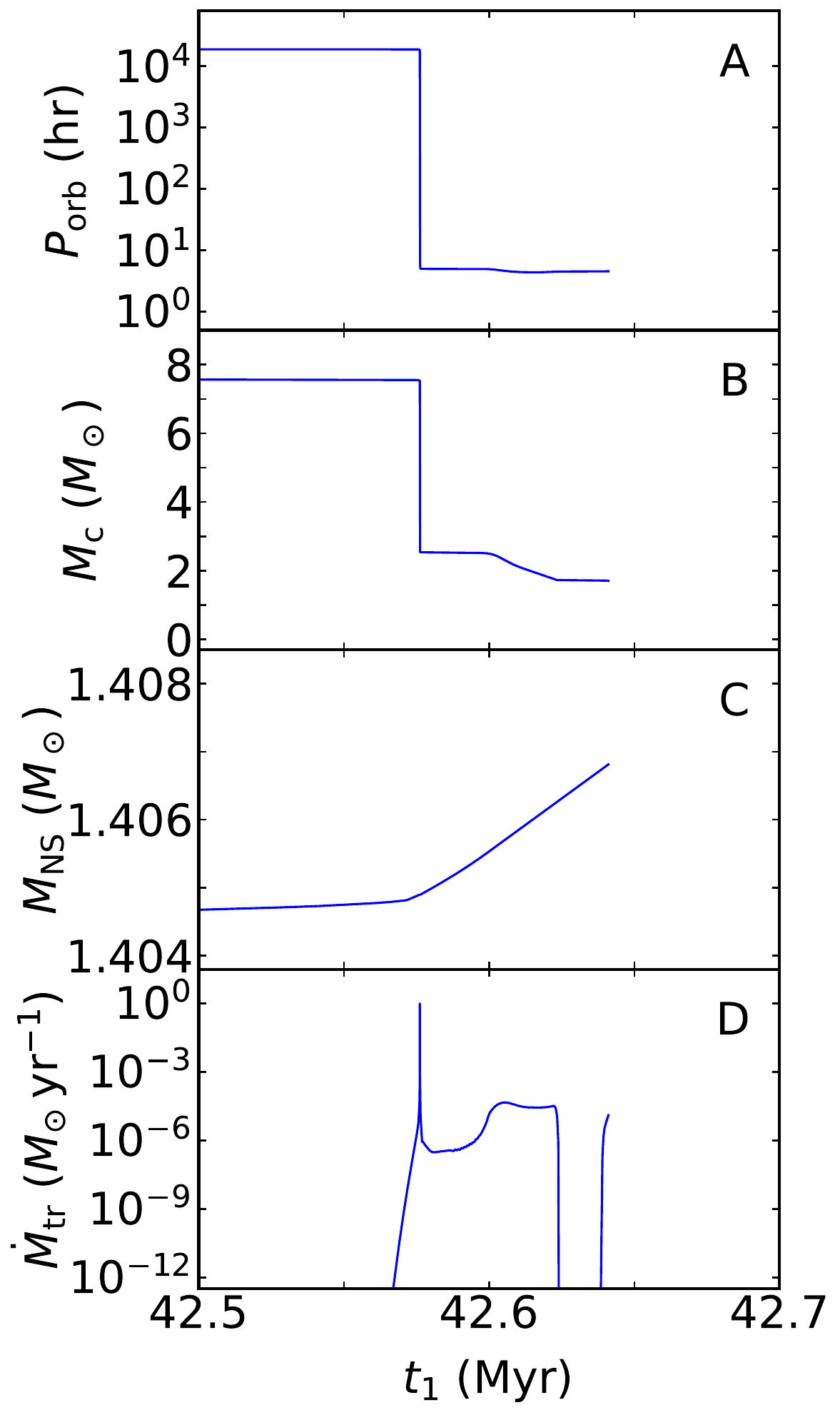}
  \includegraphics[width=0.32\textwidth]{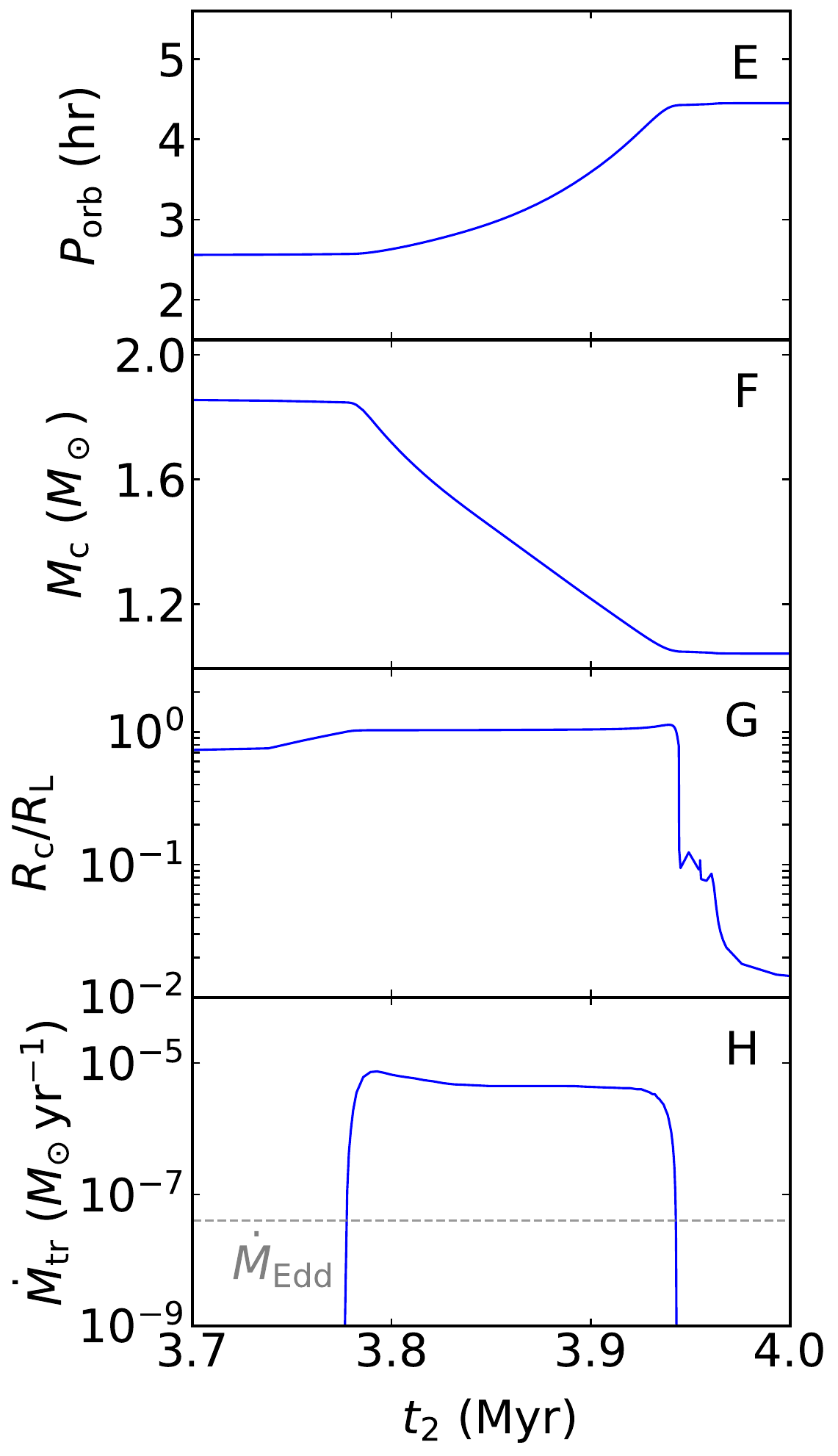}
  \includegraphics[width=0.325\textwidth]{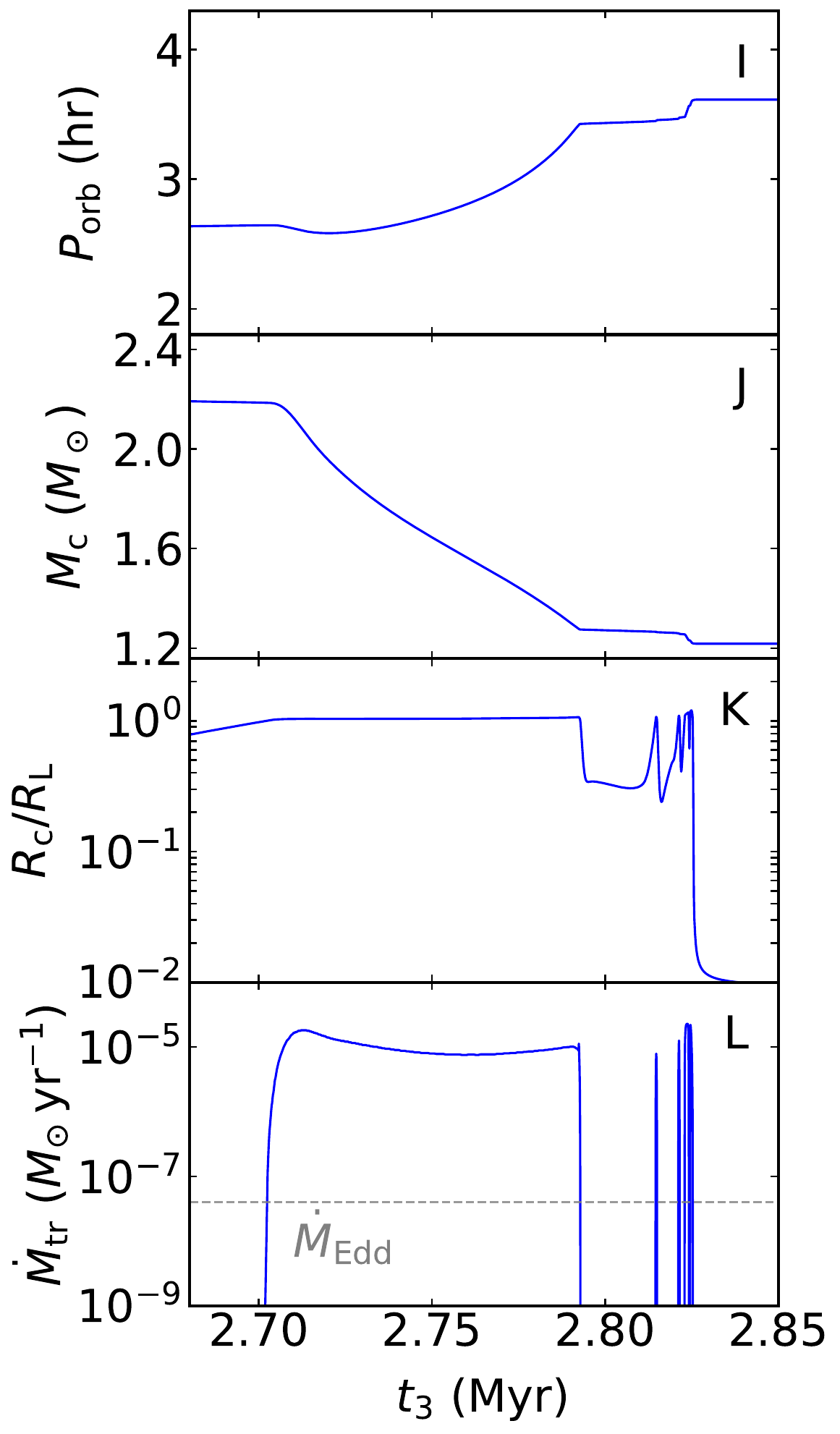}
      \caption{{\bf Other possible channels to form a PSR~J1928+1815-like binary.} 
      Results were obtained from \textsc{Mesa} simulations. 
      ({\bf A-D}): Binary evolution via Case~C around the time of mass transfer.
      The initial binary ($t_1=0$) consists of a neutron star with a mass of $1.4M_{\odot}$ and a main sequence star with a mass of $8M_{\odot}$ in a $\sim720$-day orbit, which evolves through Case~C common envelope evolution. Simulated ({\bf A}) orbital period $P_{\rm orb}$, ({\bf B}) companion star mass $M_{\rm c}$, ({\bf C}) neutron star mass ${M}_{\rm NS}$, and ({\bf D}) mass transfer rate $\dot{M}_{\rm tr}$. The Case~C RLO starts at $t_1=42.57$~million years. The system becomes a PSR~J1928+1815-like system at $t_1=42.625$~million years and survives for about $10^4$~yr, and then at $42.64$~million years the companion expands again to fill its Roche lobe. 
      ({\bf E-H}) a proto-WD companion formed via Case~BB RLO.  
      The initial neutron star binary has a helium star companion of 2.0 $M_\odot$ and an orbital period of 0.1 days at $t_2=0$ (see text). 
      Simulated ({\bf E}) $P_{\rm orb}$, ({\bf F}) $M_{\rm c}$, ({\bf G}) $R_{\rm c}/R_{\rm L}$, and ({\bf H}) $\dot{M}_{\rm tr}$. The Case~BB RLO starts at $t_2=3.78$~million years and ends at 3.94~million years, then the companion becomes a proto-WD which shrinks rapidly to $R_{\rm c}/R_{\rm L}<0.02$ in $\sim3\times10^4$~yr, with a size that can eclipse the pulsar signal. Afterward, it is too small to eclipse the pulsar signal for a large fraction of the orbit. 
      %
      ({\bf I-L}) same as (E-H) but with an initial helium star mass of 2.4 $M_\odot$. After Case~BB RLO, the companion has a thick helium envelope and a burning carbon-oxygen core at around $t_3=2.80$~million years, the produced binary system is a PSR~J1928+1815-like system, lasting for $\sim2\times10^4$~yr before a new RLO phase initiates.
         }
  \label{CE-EwT}
\end{figure}

\begin{figure}
\centering
  \includegraphics[width=0.5\textwidth]{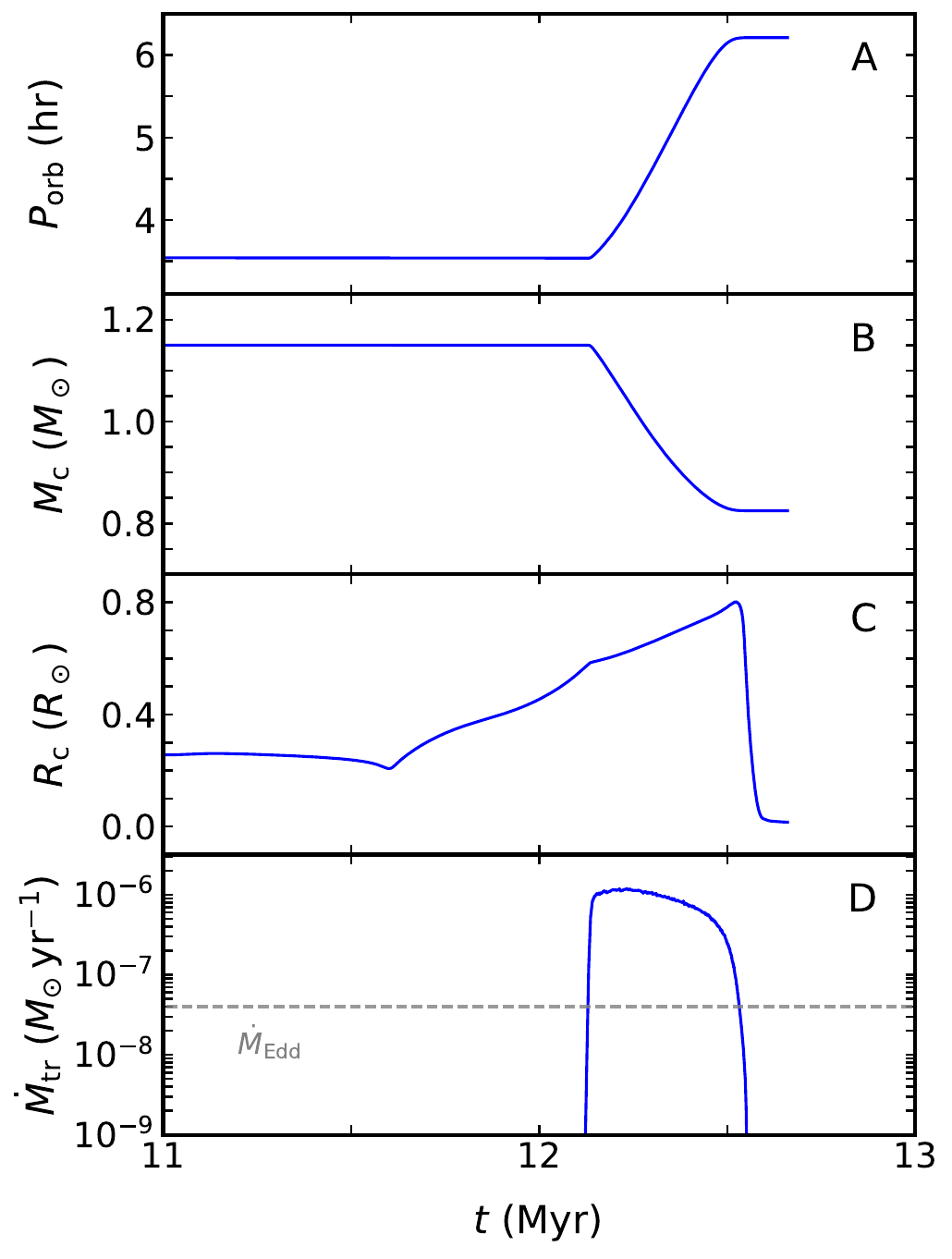}
      \caption{{\bf Simulated future evolution of the PSR J1928+1815 binary.} We simulated a binary with a helium star with a mass of 1.15 $M_\odot$ and a neutron star with a mass of 1.40 $M_\odot$ in an orbit with a period of $\rm 3.6\,hr$, and investigate its evolution in \textsc{Mesa} on ({\bf A}) the orbital period $ P_{\rm orb}$, ({\bf B}) companion mass $M_{\rm c}$, ({\bf C}) companion radius $R_{\rm c}$, and ({\bf D}) the mass-transfer rate $\dot{M}_{\rm tr}$  as functions of time $t$ (started from the current state). At about 12.1 million years the accretion exceeds the  Eddington rate,  $\rm 4\times10^{-8}\, M_\odot\,{ yr^{-1}}$,  indicated by the grey dashed line in panel~D, and the system becomes a pulsar-white dwarf binary.}
        \label{casebb_yl}
\end{figure}


\begin{table*}[ht]
\centering
\footnotesize
\caption{{\bf Log of FAST observations of PSR~J1928+1815.}  Listed are the Gregorian observation date and corresponding MJD, object name jointed with FAST beam name, FAST project number,  project principal investigator (PI), observation length in minutes, and the orbital phase range for each observation session. The start of the orbital phase is set to the time of ascending node passage. }
    \vspace{5mm}
    \begin{tabular}{c|c|l|l|c|c|c}
    \hline
     Obs. Date & MJD  & Object \& FAST Beam & Project ID & PI & Obs. Length & Orbit Phase \\
     (yyyymmdd) &     &     &    &      & (min)       &   \\
     \hline
     20200531 & 58999 & G53.48+0.51M05P3 & ZD2020\_2 & J.L. Han & 5 & 0.36 to 0.38 \\
     20200810 & 59070 & J1928+1816M01 & ZD2020\_2 & J.L. Han & 15 & 0.71 to 0.77 \\
     20200828 & 59089 & J1928+1816M01 & ZD2020\_2 & J.L. Han & 15 & 0.72 to 0.79 \\
     20200910 & 59102 & J1928+1816M01 & ZD2020\_2 & J.L. Han & 15 & 0.40 to 0.47 \\
     20201001 & 59123 & J1928+1816M01 & ZD2020\_2 & J.L. Han & 15 & 0.23 to 0.30 \\
     20201106 & 59159 & J1928+1816M01 & ZD2020\_2 & J.L. Han & 15 & 0.76 to 0.83 \\
     20201228 & 59211 & J1928+1816M01 & ZD2020\_2 & J.L. Han & 20 & 0.94 to 1.04 \\
     20201228 & 59211 & J1928+1816M01 & ZD2020\_2 & J.L. Han & 20 & 0.64 to 0.73 \\
     20201231 & 59214 & J1928+1816M01 & ZD2020\_2 & J.L. Han & 20 & 0.17 to 0.27 \\
     20201231 & 59214 & J1928+1816M01 & ZD2020\_2 & J.L. Han & 20 & 0.59 to 0.68 \\
     20210218 & 59263 & J1928+1816M01 & ZD2020\_2 & J.L. Han & 15 & 0.64 to 0.71 \\
     20210313 & 59285 & J192810+181632M01 & ZD2020\_2 & J.L. Han & 168 & 0.84 to 1.62\\
     20210313 & 59286 & J192810+181632M01 & ZD2020\_2 & J.L. Han & 14 & 0.66 to 0.73 \\
     20210405 & 59308 & J192810+181632M01 & ZD2020\_2 & J.L. Han & 240 & 0.81 to 1.93 \\
     20210713& 59407 & J192810+181632gbM01 & ZD2020\_2 & J.L. Han & 15 & 0.37 to 0.44 \\
     20210825 & 59451 & J1928+1816M01 & PT2021\_0037 & Tao Wang & 15 & 0.82 to 0.88 \\
     20211008 & 59495 & J1928+1816M01 & PT2021\_0037 & Tao Wang & 15 & 0.62 to 0.68 \\
     20211114 & 59532 & J192810+181632gbM01 & ZD2021\_2 & J.L. Han & 15 & 0.98 to 1.05 \\
     20211213 & 59561 & J1928+1816M01 & PT2021\_0037 & Tao Wang & 15 & 0.76 to 0.83 \\
     20220217 & 59627 & J1928+1816M01 & PT2021\_0037 & Tao Wang & 15 & 0.10 to 0.17 \\
     20220328 & 59666 & J1928+1816M01 & PT2021\_0037 & Tao Wang & 15 & 0.21 to 0.28 \\
     20220509 & 59707 & J1928+1816M01 & PT2021\_0037 & Tao Wang & 15 & 0.00 to 0.07 \\
     20220605 & 59734 & J1928+1816M01 & PT2021\_0037 & Tao Wang & 15 & 0.69 to 0.76 \\ 
     20230623 & 60117 & J1928+1816M01 & ZD2022\_2 & J.L. Han & 15 & 0.88 to 0.94 \\
     20231106 & 60254 & J1928+1816M01 & PT2023\_0193 & Z.L. Yang & 15 & 0.74 to 0.81 \\
     20231218 & 60296 & J1928+1816M01 & PT2023\_0193 & Z.L. Yang & 156 & 0.65 to 1.37 \\
     20241224 & 60668 & J1928+1816M01 & PT2024\_0231 & Z.L. Yang & 5 & 0.47 to 0.49 \\
    \hline
     \end{tabular}
     \label{obsinfo}
\end{table*}

\begin{table*}
\centering
\footnotesize
\caption{{\bf Estimated magnitudes of the companion, compared to the observed upper limits and estimated extinction.}  Magnitudes in each band were estimated for a naked helium star with a mass of 1.0 and 1.6~$ M_\odot$ at distance of 8 kpc by using \textsc{Mesa} (see text).  Corresponding magnitudes of the modeled stripped stars with a hydrogen envelope are obtained by fitting previous simulations with solar metallicity \cite{Gotberg+2018A&A...615A..78G}. Corresponding magnitudes for observed stripped stars with a hydrogen envelope were estimated by assuming that the Roche-lobe radii is the maximum radius of the star, while the minimum effective temperature and magnitudes are obtained from fitting the stellar properties of observed stripped stars \cite{Gotberg+2023ApJ...959..125G}. Estimated extinction magnitudes are from a 3D Milky Way dust model \cite{Marshall+2006A&A...453..635M}. The observational limits are taken from the Pan-STARRS1\cite{Chambers+2016arXiv161205560C} or UKIDSS\cite{Lawrence+2007MNRAS.379.1599L} surveys without correction for extinction. 
}
%
\vspace{5mm}
    \begin{tabular}{c|cc|cc|cc|cc}
    \hline
Parameters  &\multicolumn{2}{c|}{moldeled naked}&\multicolumn{2}{c|}{modeled stripped}&\multicolumn{2}{c|}{observed stripped}&extinction &Obs. \\
    &\multicolumn{2}{c|}{helium stars}& \multicolumn{2}{c|}{helium stars} &\multicolumn{2}{c|}{helium stars} & in the band& limits  \\
    \hline
     & &  & &  &  & & & \\[-4mm]
$M_{\rm He}$($M_{\odot}$)            & 1.0&1.6& 1.0&1.6&1.0&1.6&$\cdots$ &$\cdots$ \\
${\rm log_{10}}(L_{\rm He}/L_{\odot})$&$\cdots$ &$\cdots$ &$\cdots$ &$\cdots$& $2.41^{+0.75}_{-0.77}$&$3.11^{+0.78}_{-0.80}$& $\cdots$ &$\cdots$\\
$T_{\rm eff}$ ($\rm kK$) & 50  & 63   & 43  &  53 & $31^{+13}_{-14}$&$42^{+19}_{-19}$ &$\cdots$ & $\cdots$  \\ 
\hline
$g$ (mag)            & 17.2& 16.1 & 16.4& 15.5  &$15.81^{+0.47}_{-0.44}$ & $14.94^{+0.35}_{-0.35}$  &17.7&23.3  \\
$r$ (mag)            & 17.7& 16.6 & 16.9& 16.0  &$16.24^{+0.36}_{-0.34}$ & $15.41^{+0.27}_{-0.27}$  &13.2&23.2  \\
$i$ (mag)            & 18.1& 17.0 & 17.3& 16.4  &$16.60^{+0.30}_{-0.28}$ & $15.79^{+0.22}_{-0.22}$  &9.9 &23.1  \\
$z$ (mag)            & 18.4& 17.4 & 17.6& 16.7  &$16.91^{+0.25}_{-0.23}$ & $16.1^{+0.19}_{-0.16}$  &7.8 &22.3  \\
$y$ (mag)            & 18.7& 17.6 & 17.9& 17.0  &$17.15^{+0.22}_{-0.21}$ & $16.36^{+0.16}_{-0.16}$  &6.4 &21.4 \\
$J$ (mag)            & 18.1& 17.0 & 17.3& 16.4  &$16.56^{+0.19}_{-0.17}$ & $15.78^{+0.14}_{-0.13}$  &4.0 &19.9 \\
$H$ (mag)            & 18.2& 17.2 & 17.4& 16.5  &$16.66^{+0.14}_{-0.13}$ & $15.90^{+0.10}_{-0.10}$  &2.4 &19.0 \\
$K$ (mag)            & 18.3& 17.3 & 17.5& 16.6  &$16.75^{+0.10}_{-0.10}$ & $16.00^{+0.08}_{-0.08}$  &1.5 &18.8  \\
     \hline
     \end{tabular}    
     \label{magnitude}
\end{table*}

\begin{table*}[ht]
\centering
\footnotesize
\caption{{\bf Population synthesis calculations.} Results are from our calculations using the \textsc{Bse} population synthesis code \cite{hurl02,shao14}. Each model assumed the listed values of the common envelope ejection efficiencies $\alpha_{\rm CE}$, kick velocity dispersions $V_{\rm CC}$ of nascent neutron stars from  core-collapse supernovae, kick velocity dispersions $V_{\rm EC}$ of neutron stars from electron-capture supernovae for simulations. Also listed are the calculated formation rate of the PSR~J1928+1815-like binary systems, and the number of such  system expected in the Milky Way with a helium star companion or a proto-white dwarf companion.
}
    \vspace{5mm}
    \begin{tabular}{ccccccc}
    \hline
    Model& $\alpha_{\rm CE}$  & $V_{\rm CC}$&$V_{\rm EC}$  & Formation rate & Binary Number & Binary Number \\
      &   & ($\rm km\,s^{-1}$) & ($\rm km\,s^{-1}$)    &  ($\rm 10^{-6} yr^{-1}$)          &    helium star  &  proto-white dwarf   \\
     \hline
  A &    0.3   & 265 &  40    &   3.4   &   41 & 0.017  \\
  B &    0.3   & 265 &  80    &   1.5   &   19 & 0.007  \\
  C &    0.3   & 320 &  80    &   1.3   &   16 & 0.007  \\
  D &    1.0   & 265 &  40    &   7.2  &  84 & 0.035   \\
  E &    1.0   & 265 &  80    &   3.8  &  45 & 0.019   \\
  F &    1.0   & 320 &  80    &   3.3  &  39 & 0.016   \\
  G &    3.0   & 265 &  40    &   5.8   &   62 & 0.030  \\
  H &    3.0   & 265 &  80    &   3.3   &   38 & 0.017  \\
  I &    3.0   & 320 &  80    &  3.0   &   33 & 0.015  \\
        \hline
     \end{tabular}
     \label{birthrate and number}
\end{table*}

\clearpage


\begin{thebibliography}{100}
\providecommand{\url}[1]{\texttt{#1}}
\expandafter\ifx\csname urlstyle\endcsname\relax
  \providecommand{\doi}[1]{doi:\discretionary{}{}{}#1}\else
  \providecommand{\doi}{doi:\discretionary{}{}{}\begingroup \urlstyle{rm}\Url}\fi

\bibitem{Alpar+1982Natur.300..728A}
M.~A. {Alpar}, A.~F. {Cheng}, M.~A. {Ruderman}, J.~{Shaham}, {A new class of radio pulsars}. \emph{\nat} \textbf{300}~(5894), 728--730 (1982), \doi{10.1038/300728a0}.

\bibitem{Han+2020RAA....20..161H}
Z.-W. {Han}, H.-W. {Ge}, X.-F. {Chen}, H.-L. {Chen}, {Binary Population Synthesis}. \emph{Res. Astron. Astrophys.} \textbf{20}~(10), 161 (2020), \doi{10.1088/1674-4527/20/10/161}.

\bibitem{Tauris+2023pbse.book.....T}
T.~M. Tauris, E.~P.~J. van~den Heuvel, \emph{Physics of Binary Star Evolution: From Stars to X-ray Binaries and Gravitational Wave Sources}, vol.~68 of \emph{Princeton Series in Astrophysics} (Princeton University Press) (2023), \url{http://www.jstor.org/stable/jj.8595663}.

\bibitem{Paczynski+1976IAUS...73...75P}
B.~{Paczynski}, {Common Envelope Binaries}, in \emph{Structure and Evolution of Close Binary Systems}, P.~{Eggleton}, S.~{Mitton}, J.~{Whelan}, Eds., vol.~73 of \emph{IAU Symposium} (1976), p.~75.

\bibitem{Iben+1993PASP..105.1373I}
J.~{Iben}, Icko, M.~{Livio}, {Common Envelopes in Binary Star Evolution}. \emph{\pasp} \textbf{105}, 1373 (1993), \doi{10.1086/133321}.

\bibitem{Ivanova+2013A&ARv..21...59I}
N.~{Ivanova}, \emph{et~al.}, {Common envelope evolution: where we stand and how we can move forward}. \emph{\aapr} \textbf{21}, 59 (2013), \doi{10.1007/s00159-013-0059-2}.

\bibitem{Drout+2023Sci...382.1287D}
M.~R. {Drout}, \emph{et~al.}, {An observed population of intermediate-mass helium stars that have been stripped in binaries}. \emph{Sci} \textbf{382}~(6676), 1287--1291 (2023), \doi{10.1126/science.ade4970}.

\bibitem{Chen_2013}
H.-L. {Chen}, X.~{Chen}, T.~M. {Tauris}, Z.~{Han}, {Formation of Black Widows and Redbacks{\textemdash}Two Distinct Populations of Eclipsing Binary Millisecond Pulsars}. \emph{\apj} \textbf{775}~(1), 27 (2013), \doi{10.1088/0004-637X/775/1/27}.

\bibitem{Pan+2021ApJ...915L..28P}
Z.~{Pan}, \emph{et~al.}, {FAST Globular Cluster Pulsar Survey: Twenty-four Pulsars Discovered in 15 Globular Clusters}. \emph{\apjl} \textbf{915}~(2), L28 (2021), \doi{10.3847/2041-8213/ac0bbd}.

\bibitem{Roberts+2013IAUS..291..127R}
M.~S.~E. {Roberts}, {Surrounded by spiders! New black widows and redbacks in the Galactic field}, in \emph{Neutron Stars and Pulsars: Challenges and Opportunities after 80 years}, J.~{van Leeuwen}, Ed., vol. 291 of \emph{IAU Symposium} (2013), pp. 127--132, \doi{10.1017/S174392131202337X}.

\bibitem{Smith+2023ApJ...958..191S}
D.~A. {Smith}, \emph{et~al.}, {The Third Fermi Large Area Telescope Catalog of Gamma-Ray Pulsars}. \emph{\apj} \textbf{958}~(2), 191 (2023), \doi{10.3847/1538-4357/acee67}.

\bibitem{Nan+2006ScChG..49..129N}
R.~{Nan}, {Five hundred meter aperture spherical radio telescope (FAST)}. \emph{Science in China: Physics, Mechanics and Astronomy} \textbf{49}~(2), 129--148 (2006), \doi{10.1007/s11433-006-0129-9}.

\bibitem{Han+2021RAA....21..107H}
J.~L. {Han}, \emph{et~al.}, {The FAST Galactic Plane Pulsar Snapshot survey: I. Project design and pulsar discoveries}. \emph{Res. Astron. Astrophys.} \textbf{21}~(5), 107 (2021), \doi{10.1088/1674-4527/21/5/107}.

\bibitem{mm}
Materials and methods are available as supplementary materials.

\bibitem{Jiang+2020RAA....20...64J}
P.~{Jiang}, \emph{et~al.}, {The fundamental performance of FAST with 19-beam receiver at L band}. \emph{Res. Astron. Astrophys.} \textbf{20}~(5), 064 (2020), \doi{10.1088/1674-4527/20/5/64}.

\bibitem{Eggleton+1983ApJ...268..368E}
P.~P. {Eggleton}, {Aproximations to the radii of Roche lobes.} \emph{\apj} \textbf{268}, 368--369 (1983), \doi{10.1086/160960}.

\bibitem{Demircan+1991Ap&SS.181..313D}
O.~{Demircan}, G.~{Kahraman}, {Stellar Mass / Luminosity and Mass / Radius Relations}. \emph{\apss} \textbf{181}~(2), 313--322 (1991), \doi{10.1007/BF00639097}.

\bibitem{Boyajian+2012ApJ...757..112B}
T.~S. {Boyajian}, \emph{et~al.}, {Stellar Diameters and Temperatures. II. Main-sequence K- and M-stars}. \emph{\apj} \textbf{757}~(2), 112 (2012), \doi{10.1088/0004-637X/757/2/112}.

\bibitem{Thompson+1994ApJ...422..304T}
C.~{Thompson}, R.~D. {Blandford}, C.~R. {Evans}, E.~S. {Phinney}, {Physical Processes in Eclipsing Pulsars: Eclipse Mechanisms and Diagnostics}. \emph{\apj} \textbf{422}, 304 (1994), \doi{10.1086/173728}.

\bibitem{Wadiasingh+2017ApJ...839...80W}
Z.~{Wadiasingh}, A.~K. {Harding}, C.~{Venter}, M.~{B{\"o}ttcher}, M.~G. {Baring}, {Constraining Relativistic Bow Shock Properties in Rotation-powered Millisecond Pulsar Binaries}. \emph{\apj} \textbf{839}~(2), 80 (2017), \doi{10.3847/1538-4357/aa69bf}.

\bibitem{Du+2023RAA....23l5024D}
Z.-X. {Du}, \emph{et~al.}, {Constraining the Orbital Inclination and Companion Properties of Three Black Widow Pulsars Detected by FAST}. \emph{Res. Astron. Astrophys.} \textbf{23}~(12), 125024 (2023), \doi{10.1088/1674-4527/ad034b}.

\bibitem{Cordes+2002astro.ph..7156C}
J.~M. {Cordes}, T.~J.~W. {Lazio}, {NE2001.I. A New Model for the Galactic Distribution of Free Electrons and its Fluctuations}. \emph{arXiv e-prints} astro-ph/0207156 (2002).

\bibitem{Yao+2017ApJ...835...29Y}
J.~M. {Yao}, R.~N. {Manchester}, N.~{Wang}, {A New Electron-density Model for Estimation of Pulsar and FRB Distances}. \emph{\apj} \textbf{835}~(1), 29 (2017), \doi{10.3847/1538-4357/835/1/29}.

\bibitem{Lawrence+2007MNRAS.379.1599L}
A.~{Lawrence}, \emph{et~al.}, {The UKIRT Infrared Deep Sky Survey (UKIDSS)}. \emph{\mnras} \textbf{379}~(4), 1599--1617 (2007), \doi{10.1111/j.1365-2966.2007.12040.x}.

\bibitem{Koljonen+2023MNRAS.525.3963K}
K.~I.~I. {Koljonen}, M.~{Linares}, {A Gaia view of the optical and X-ray luminosities of compact binary millisecond pulsars}. \emph{\mnras} \textbf{525}~(3), 3963--3985 (2023), \doi{10.1093/mnras/stad2485}.

\bibitem{Boller+2016A&A...588A.103B}
T.~{Boller}, \emph{et~al.}, {Second ROSAT all-sky survey (2RXS) source catalogue}. \emph{\aap} \textbf{588}, A103 (2016), \doi{10.1051/0004-6361/201525648}.

\bibitem{Evans+2020ApJS..247...54E}
P.~A. {Evans}, \emph{et~al.}, {2SXPS: An Improved and Expanded Swift X-Ray Telescope Point-source Catalog}. \emph{\apjs} \textbf{247}~(2), 54 (2020), \doi{10.3847/1538-4365/ab7db9}.

\bibitem{Abdollahi+2022ApJS..260...53A}
S.~{Abdollahi}, \emph{et~al.}, {Incremental Fermi Large Area Telescope Fourth Source Catalog}. \emph{\apjs} \textbf{260}~(2), 53 (2022), \doi{10.3847/1538-4365/ac6751}.

\bibitem{Lazarus+2014MNRAS.437.1485L}
P.~{Lazarus}, \emph{et~al.}, {Timing of a young mildly recycled pulsar with a massive white dwarf companion}. \emph{\mnras} \textbf{437}~(2), 1485--1494 (2014), \doi{10.1093/mnras/stt1996}.

\bibitem{Tauris+2013A&A...558A..39T}
T.~M. {Tauris}, D.~{Sanyal}, S.~C. {Yoon}, N.~{Langer}, {Evolution towards and beyond accretion-induced collapse of massive white dwarfs and formation of millisecond pulsars}. \emph{\aap} \textbf{558}, A39 (2013), \doi{10.1051/0004-6361/201321662}.

\bibitem{Radhakrishnan+1982CSci...51.1096R}
V.~{Radhakrishnan}, G.~{Srinivasan}, {On the origin of the recently discovered ultra-rapid pulsar}. \emph{Curr. Sci.} \textbf{51}, 1096--1099 (1982).

\bibitem{Ge+2015ApJ...812...40G}
H.~{Ge}, R.~F. {Webbink}, X.~{Chen}, Z.~{Han}, {Adiabatic Mass Loss in Binary Stars. II. From Zero-age Main Sequence to the Base of the Giant Branch}. \emph{\apj} \textbf{812}~(1), 40 (2015), \doi{10.1088/0004-637X/812/1/40}.

\bibitem{Ge+2020ApJ...899..132G}
H.~{Ge}, R.~F. {Webbink}, X.~{Chen}, Z.~{Han}, {Adiabatic Mass Loss in Binary Stars. III. From the Base of the Red Giant Branch to the Tip of the Asymptotic Giant Branch}. \emph{\apj} \textbf{899}~(2), 132 (2020), \doi{10.3847/1538-4357/aba7b7}.

\bibitem{Tauris+2011MNRAS.416.2130T}
T.~M. {Tauris}, N.~{Langer}, M.~{Kramer}, {Formation of millisecond pulsars with CO white dwarf companions - I. PSR J1614-2230: evidence for a neutron star born massive}. \emph{\mnras} \textbf{416}~(3), 2130--2142 (2011), \doi{10.1111/j.1365-2966.2011.19189.x}.

\bibitem{Dewi+2002MNRAS.331.1027D}
J.~D.~M. {Dewi}, O.~R. {Pols}, G.~J. {Savonije}, E.~P.~J. {van den Heuvel}, {The evolution of naked helium stars with a neutron star companion in close binary systems}. \emph{\mnras} \textbf{331}~(4), 1027--1040 (2002), \doi{10.1046/j.1365-8711.2002.05257.x}.

\bibitem{Houck+1991ApJ...376..234H}
J.~C. {Houck}, R.~A. {Chevalier}, {Steady Spherical Hypercritical Accretion onto Neutron Stars}. \emph{\apj} \textbf{376}, 234 (1991), \doi{10.1086/170272}.

\bibitem{Bhattacharya+1991PhR...203....1B}
D.~{Bhattacharya}, E.~P.~J. {van den Heuvel}, {Formation and evolution of binary and millisecond radio pulsars}. \emph{\physrep} \textbf{203}~(1-2), 1--124 (1991), \doi{10.1016/0370-1573(91)90064-S}.

\bibitem{Krolik+1991ApJ...373L..69K}
J.~H. {Krolik}, {Multipolar Magnetic Fields in Neutron Stars}. \emph{\apjl} \textbf{373}, L69 (1991), \doi{10.1086/186053}.

\bibitem{Radhakrishnan+1969ApL.....3..225R}
V.~{Radhakrishnan}, D.~J. {Cooke}, {Magnetic Poles and the Polarization Structure of Pulsar Radiation}. \emph{\aplett} \textbf{3}, 225 (1969).

\bibitem{Biryukov+2021MNRAS.505.1775B}
A.~{Biryukov}, P.~{Abolmasov}, {Magnetic angle evolution in accreting neutron stars}. \emph{\mnras} \textbf{505}~(2), 1775--1786 (2021), \doi{10.1093/mnras/stab1378}.

\bibitem{Pulsar_Astronomy2022}
A.~Lyne, F.~Graham-Smith, B.~Stappers, \emph{Pulsar Astronomy}, Cambridge Astrophysics (Cambridge University Press), 5 ed. (2022).

\bibitem{Gotberg+2023ApJ...959..125G}
Y.~{G{\"o}tberg}, \emph{et~al.}, {Stellar Properties of Observed Stars Stripped in Binaries in the Magellanic Clouds}. \emph{\apj} \textbf{959}~(2), 125 (2023), \doi{10.3847/1538-4357/ace5a3}.

\bibitem{Thorne+1975ApJ...199L..19T}
K.~S. {Thorne}, A.~N. {Zytkow}, {Red giants and supergiants with degenerate neutron cores.} \emph{\apjl} \textbf{199}, L19--L24 (1975), \doi{10.1086/181839}.

\bibitem{Thorne+1977ApJ...212..832T}
K.~S. {Thorne}, A.~N. {Zytkow}, {Stars with degenerate neutron cores. I. Structure of equilibrium models.} \emph{\apj} \textbf{212}, 832--858 (1977), \doi{10.1086/155109}.

\bibitem{Hobbs+2006MNRAS.369..655H}
G.~B. {Hobbs}, R.~T. {Edwards}, R.~N. {Manchester}, {TEMPO2, a new pulsar-timing package - I. An overview}. \emph{\mnras} \textbf{369}~(2), 655--672 (2006), \doi{10.1111/j.1365-2966.2006.10302.x}.

\bibitem{Ransom+2001PhDT.......123R}
S.~M. {Ransom}, \emph{{New search techniques for binary pulsars}}, Ph.D. thesis, Harvard University, Massachusetts (2001).

\bibitem{Freire+2001MNRAS.322..885F}
P.~C. {Freire}, M.~{Kramer}, A.~G. {Lyne}, {Determination of the orbital parameters of binary pulsars}. \emph{\mnras} \textbf{322}~(4), 885--890 (2001), \doi{10.1046/j.1365-8711.2001.04200.x}.

\bibitem{Bhattacharyya+2008MNRAS.387..273B}
B.~{Bhattacharyya}, R.~{Nityananda}, {Determination of the orbital parameters of binary pulsars}. \emph{\mnras} \textbf{387}~(1), 273--278 (2008), \doi{10.1111/j.1365-2966.2008.13213.x}.

\bibitem{Straten+2011PASA...28....1V}
W.~{van Straten}, M.~{Bailes}, {DSPSR: Digital Signal Processing Software for Pulsar Astronomy}. \emph{\pasa} \textbf{28}~(1), 1--14 (2011), \doi{10.1071/AS10021}.

\bibitem{Hotan+2004PASA...21..302H}
A.~W. {Hotan}, W.~{van Straten}, R.~N. {Manchester}, {PSRCHIVE and PSRFITS: An Open Approach to Radio Pulsar Data Storage and Analysis}. \emph{\pasa} \textbf{21}~(3), 302--309 (2004), \doi{10.1071/AS04022}.

\bibitem{Chen+2023RAA....23j4004C}
X.~{Chen}, J.~L. {Han}, W.~Q. {Su}, Z.~L. {Yang}, D.~J. {Zhou}, {Cleaning Radio Frequency Interference in Pulsar-Folded Data Based on the Conditional Random Fields with an Adaptive Prior}. \emph{Res. Astron. Astrophys.} \textbf{23}~(10), 104004 (2023), \doi{10.1088/1674-4527/acd52b}.

\bibitem{DE440}
R.~S. {Park}, W.~M. {Folkner}, J.~G. {Williams}, D.~H. {Boggs}, {The JPL Planetary and Lunar Ephemerides DE440 and DE441}. \emph{\aj} \textbf{161}~(3), 105 (2021), \doi{10.3847/1538-3881/abd414}.

\bibitem{Lange+2001MNRAS.326..274L}
C.~{Lange}, \emph{et~al.}, {Precision timing measurements of PSR J1012+5307}. \emph{\mnras} \textbf{326}~(1), 274--282 (2001), \doi{10.1046/j.1365-8711.2001.04606.x}.

\bibitem{g2022J1952}
T.~{Gautam}, \emph{et~al.}, {Relativistic effects in a mildly recycled pulsar binary: PSR J1952+2630}. \emph{\aap} \textbf{668}, A187 (2022), \doi{10.1051/0004-6361/202244699}.

\bibitem{vizier}
F.~{Ochsenbein}, P.~{Bauer}, J.~{Marcout}, {The VizieR database of astronomical catalogues}. \emph{\aaps} \textbf{143}, 23--32 (2000), \doi{10.1051/aas:2000169}.

\bibitem{Skrutskie+2006AJ....131.1163S}
M.~F. {Skrutskie}, \emph{et~al.}, {The Two Micron All Sky Survey (2MASS)}. \emph{\aj} \textbf{131}~(2), 1163--1183 (2006), \doi{10.1086/498708}.

\bibitem{Benjamin+2003PASP..115..953B}
R.~A. {Benjamin}, \emph{et~al.}, {GLIMPSE. I. An SIRTF Legacy Project to Map the Inner Galaxy}. \emph{\pasp} \textbf{115}~(810), 953--964 (2003), \doi{10.1086/376696}.

\bibitem{Gaia+2021A&A...649A...1G}
{Gaia Collaboration}, \emph{et~al.}, {Gaia Early Data Release 3. Summary of the contents and survey properties}. \emph{\aap} \textbf{649}, A1 (2021), \doi{10.1051/0004-6361/202039657}.

\bibitem{Wright+2010AJ....140.1868W}
E.~L. {Wright}, \emph{et~al.}, {The Wide-field Infrared Survey Explorer (WISE): Mission Description and Initial On-orbit Performance}. \emph{\aj} \textbf{140}~(6), 1868--1881 (2010), \doi{10.1088/0004-6256/140/6/1868}.

\bibitem{Mainzer+2011ApJ...731...53M}
A.~{Mainzer}, \emph{et~al.}, {Preliminary Results from NEOWISE: An Enhancement to the Wide-field Infrared Survey Explorer for Solar System Science}. \emph{\apj} \textbf{731}~(1), 53 (2011), \doi{10.1088/0004-637X/731/1/53}.

\bibitem{Chambers+2016arXiv161205560C}
K.~C. {Chambers}, \emph{et~al.}, {The Pan-STARRS1 Surveys}. \emph{arXiv e-prints} arXiv:1612.05560 (2016), \doi{10.48550/arXiv.1612.05560}.

\bibitem{2007A&A...467..777C}
M.~{Casali}, \emph{et~al.}, {The UKIRT wide-field camera}. \emph{\aap} \textbf{467}~(2), 777--784 (2007), \doi{10.1051/0004-6361:20066514}.

\bibitem{2006MNRAS.367..454H}
P.~C. {Hewett}, S.~J. {Warren}, S.~K. {Leggett}, S.~T. {Hodgkin}, {The UKIRT Infrared Deep Sky Survey ZY JHK photometric system: passbands and synthetic colours}. \emph{\mnras} \textbf{367}~(2), 454--468 (2006), \doi{10.1111/j.1365-2966.2005.09969.x}.

\bibitem{Hambly+2008MNRAS.384..637H}
N.~C. {Hambly}, \emph{et~al.}, {The WFCAM Science Archive}. \emph{\mnras} \textbf{384}~(2), 637--662 (2008), \doi{10.1111/j.1365-2966.2007.12700.x}.

\bibitem{1988Natur.333..237F}
A.~S. {Fruchter}, D.~R. {Stinebring}, J.~H. {Taylor}, {A millisecond pulsar in an eclipsing binary}. \emph{\nat} \textbf{333}~(6170), 237--239 (1988), \doi{10.1038/333237a0}.

\bibitem{Vink+2017A&A...607L...8V}
J.~S. {Vink}, {Winds from stripped low-mass helium stars and Wolf-Rayet stars}. \emph{\aap} \textbf{607}, L8 (2017), \doi{10.1051/0004-6361/201731902}.

\bibitem{Gotberg+2018A&A...615A..78G}
Y.~{G{\"o}tberg}, \emph{et~al.}, {Spectral models for binary products: Unifying subdwarfs and Wolf-Rayet stars as a sequence of stripped-envelope stars}. \emph{\aap} \textbf{615}, A78 (2018), \doi{10.1051/0004-6361/201732274}.

\bibitem{Canto+1996ApJ...469..729C}
J.~{Cant{\'o}}, A.~C. {Raga}, F.~P. {Wilkin}, {Exact, Algebraic Solutions of the Thin-Shell Two-Wind Interaction Problem}. \emph{\apj} \textbf{469}, 729 (1996), \doi{10.1086/177820}.

\bibitem{2011ApJS..192....3P}
B.~{Paxton}, \emph{et~al.}, {Modules for Experiments in Stellar Astrophysics (MESA)}. \emph{\apjs} \textbf{192}~(1), 3 (2011), \doi{10.1088/0067-0049/192/1/3}.

\bibitem{Paxton+2019ApJS..243...10P}
B.~{Paxton}, \emph{et~al.}, {Modules for Experiments in Stellar Astrophysics (MESA): Pulsating Variable Stars, Rotation, Convective Boundaries, and Energy Conservation}. \emph{\apjs} \textbf{243}~(1), 10 (2019), \doi{10.3847/1538-4365/ab2241}.

\bibitem{fuku96}
M.~{Fukugita}, \emph{et~al.}, {The Sloan Digital Sky Survey Photometric System}. \emph{\aj} \textbf{111}, 1748 (1996), \doi{10.1086/117915}.

\bibitem{bess98}
M.~S. {Bessell}, F.~{Castelli}, B.~{Plez}, {Model atmospheres broad-band colors, bolometric corrections and temperature calibrations for O - M stars}. \emph{\aap} \textbf{333}, 231--250 (1998).

\bibitem{Marshall+2006A&A...453..635M}
D.~J. {Marshall}, A.~C. {Robin}, C.~{Reyl{\'e}}, M.~{Schultheis}, S.~{Picaud}, {Modelling the Galactic interstellar extinction distribution in three dimensions}. \emph{\aap} \textbf{453}~(2), 635--651 (2006), \doi{10.1051/0004-6361:20053842}.

\bibitem{Lu+2013ApJ...768..193L}
G.~{L{\"u}}, C.~{Zhu}, P.~{Podsiadlowski}, {Dust Formation in the Ejecta of Common Envelope Systems}. \emph{\apj} \textbf{768}~(2), 193 (2013), \doi{10.1088/0004-637X/768/2/193}.

\bibitem{Iaconi+2020MNRAS.497.3166I}
R.~{Iaconi}, K.~{Maeda}, T.~{Nozawa}, O.~{De Marco}, T.~{Reichardt}, {Properties of the post in-spiral common envelope ejecta II: dust formation}. \emph{\mnras} \textbf{497}~(3), 3166--3179 (2020), \doi{10.1093/mnras/staa2169}.

\bibitem{Ge+2010ApJ...717..724G}
H.~{Ge}, M.~S. {Hjellming}, R.~F. {Webbink}, X.~{Chen}, Z.~{Han}, {Adiabatic Mass Loss in Binary Stars. I. Computational Method}. \emph{\apj} \textbf{717}~(2), 724--738 (2010), \doi{10.1088/0004-637X/717/2/724}.

\bibitem{Ge+2020ApJS..249....9G}
H.~{Ge}, R.~F. {Webbink}, Z.~{Han}, {The Thermal Equilibrium Mass-loss Model and Its Applications in Binary Evolution}. \emph{\apjs} \textbf{249}~(1), 9 (2020), \doi{10.3847/1538-4365/ab98f6}.

\bibitem{Ge+2023ApJ...945....7G}
H.~{Ge}, \emph{et~al.}, {Criteria for Dynamical Timescale Mass Transfer of Metal-poor Intermediate-mass Stars}. \emph{\apj} \textbf{945}~(1), 7 (2023), \doi{10.3847/1538-4357/acb7e9}.

\bibitem{hurl02}
J.~R. {Hurley}, C.~A. {Tout}, O.~R. {Pols}, {Evolution of binary stars and the effect of tides on binary populations}. \emph{\mnras} \textbf{329}~(4), 897--928 (2002), \doi{10.1046/j.1365-8711.2002.05038.x}.

\bibitem{shao14}
Y.~{Shao}, X.-D. {Li}, {On the Formation of Be Stars through Binary Interaction}. \emph{\apj} \textbf{796}~(1), 37 (2014), \doi{10.1088/0004-637X/796/1/37}.

\bibitem{Shao+2019ApJ...886..118S}
Y.~{Shao}, X.-D. {Li}, Z.-G. {Dai}, {A Population of Neutron Star Ultraluminous X-Ray Sources with a Helium Star Companion}. \emph{\apj} \textbf{886}~(2), 118 (2019), \doi{10.3847/1538-4357/ab4d50}.

\bibitem{Kroupa+1993MNRAS.262..545K}
P.~{Kroupa}, C.~A. {Tout}, G.~{Gilmore}, {The Distribution of Low-Mass Stars in the Galactic Disc}. \emph{\mnras} \textbf{262}, 545--587 (1993), \doi{10.1093/mnras/262.3.545}.

\bibitem{Kobulnicky+2007ApJ...670..747K}
H.~A. {Kobulnicky}, C.~L. {Fryer}, {A New Look at the Binary Characteristics of Massive Stars}. \emph{\apj} \textbf{670}~(1), 747--765 (2007), \doi{10.1086/522073}.

\bibitem{Abt+1983ARA&A..21..343A}
H.~A. {Abt}, {Normal and abnormal binary frequencies.} \emph{\araa} \textbf{21}, 343--372 (1983), \doi{10.1146/annurev.aa.21.090183.002015}.

\bibitem{Hobbs+2005MNRAS.360..974H}
G.~{Hobbs}, D.~R. {Lorimer}, A.~G. {Lyne}, M.~{Kramer}, {A statistical study of 233 pulsar proper motions}. \emph{\mnras} \textbf{360}~(3), 974--992 (2005), \doi{10.1111/j.1365-2966.2005.09087.x}.

\bibitem{Igoshev2020MNRAS.494.3663I}
A.~P. {Igoshev}, {The observed velocity distribution of young pulsars - II. Analysis of complete PSR{\ensuremath{\pi}}}. \emph{\mnras} \textbf{494}~(3), 3663--3674 (2020), \doi{10.1093/mnras/staa958}.

\bibitem{Verbunt+2017A&A...608A..57V}
F.~{Verbunt}, A.~{Igoshev}, E.~{Cator}, {The observed velocity distribution of young pulsars}. \emph{\aap} \textbf{608}, A57 (2017), \doi{10.1051/0004-6361/201731518}.

\bibitem{Webbink+1984ApJ...277..355W}
R.~F. {Webbink}, {Double white dwarfs as progenitors of R Coronae Borealis stars and type I supernovae.} \emph{\apj} \textbf{277}, 355--360 (1984), \doi{10.1086/161701}.

\bibitem{kool90}
M.~{de Kool}, {Common Envelope Evolution and Double Cores of Planetary Nebulae}. \emph{\apj} \textbf{358}, 189 (1990), \doi{10.1086/168974}.

\bibitem{Ge+2022ApJ...933..137G}
H.~{Ge}, \emph{et~al.}, {The Common Envelope Evolution Outcome-A Case Study on Hot Subdwarf B Stars}. \emph{\apj} \textbf{933}~(2), 137 (2022), \doi{10.3847/1538-4357/ac75d3}.

\bibitem{Ge+2024ApJ...961..202G}
H.~{Ge}, \emph{et~al.}, {The Common Envelope Evolution Outcome. II. Short-orbital-period Hot Subdwarf B Binaries Reveal a Clear Picture}. \emph{\apj} \textbf{961}~(2), 202 (2024), \doi{10.3847/1538-4357/ad158e}.

\bibitem{Ricker+2012ApJ...746...74R}
P.~M. {Ricker}, R.~E. {Taam}, {An AMR Study of the Common-envelope Phase of Binary Evolution}. \emph{\apj} \textbf{746}~(1), 74 (2012), \doi{10.1088/0004-637X/746/1/74}.

\bibitem{Andrews+2015ApJ...801...32A}
J.~J. {Andrews}, W.~M. {Farr}, V.~{Kalogera}, B.~{Willems}, {Evolutionary Channels for the Formation of Double Neutron Stars}. \emph{\apj} \textbf{801}~(1), 32 (2015), \doi{10.1088/0004-637X/801/1/32}.

\bibitem{Iben+1984ApJS...54..335I}
J.~{Iben}, I., A.~V. {Tutukov}, {Supernovae of type I as end products of the evolution of binaries with components of moderate initial mass.} \emph{\apjs} \textbf{54}, 335--372 (1984), \doi{10.1086/190932}.

\bibitem{Livio+1988ApJ...329..764L}
M.~{Livio}, N.~{Soker}, {The Common Envelope Phase in the Evolution of Binary Stars}. \emph{\apj} \textbf{329}, 764 (1988), \doi{10.1086/166419}.

\bibitem{2021A&A...650A.107M}
P.~{Marchant}, \emph{et~al.}, {The role of mass transfer and common envelope evolution in the formation of merging binary black holes}. \emph{\aap} \textbf{650}, A107 (2021), \doi{10.1051/0004-6361/202039992}.

\bibitem{2019ApJ...883L..45F}
T.~{Fragos}, \emph{et~al.}, {The Complete Evolution of a Neutron-star Binary through a Common Envelope Phase Using 1D Hydrodynamic Simulations}. \emph{\apjl} \textbf{883}~(2), L45 (2019), \doi{10.3847/2041-8213/ab40d1}.

\bibitem{jiang2021}
L.~{Jiang}, T.~M. {Tauris}, W.-C. {Chen}, J.~{Fuller}, {Novel Model of an Ultra-stripped Supernova Progenitor of a Double Neutron Star}. \emph{\apj} \textbf{920}~(2), L36 (2021), \doi{10.3847/2041-8213/ac2cc9}.

\bibitem{Guo+2024MNRAS.530.4461G}
Y.-L. {Guo}, \emph{et~al.}, {Electron-capture supernovae in NS + He star systems and the double neutron star systems}. \emph{\mnras} \textbf{530}~(4), 4461--4473 (2024), \doi{10.1093/mnras/stae1112}.

\bibitem{Kaplan+2013ApJ...765..158K}
D.~L. {Kaplan}, \emph{et~al.}, {A Metal-rich Low-gravity Companion to a Massive Millisecond Pulsar}. \emph{\apj} \textbf{765}~(2), 158 (2013), \doi{10.1088/0004-637X/765/2/158}.

\bibitem{Istrate+2014A&A...571L...3I}
A.~G. {Istrate}, T.~M. {Tauris}, N.~{Langer}, J.~{Antoniadis}, {The timescale of low-mass proto-helium white dwarf evolution}. \emph{\aap} \textbf{571}, L3 (2014), \doi{10.1051/0004-6361/201424681}.

\bibitem{Tauris+2012MNRAS.425.1601T}
T.~M. {Tauris}, N.~{Langer}, M.~{Kramer}, {Formation of millisecond pulsars with CO white dwarf companions - II. Accretion, spin-up, true ages and comparison to MSPs with He white dwarf companions}. \emph{\mnras} \textbf{425}~(3), 1601--1627 (2012), \doi{10.1111/j.1365-2966.2012.21446.x}.

\bibitem{Wang+2021MNRAS.506.4654W}
B.~{Wang}, \emph{et~al.}, {Ultracompact X-ray binaries with He star companions}. \emph{\mnras} \textbf{506}~(3), 4654--4666 (2021), \doi{10.1093/mnras/stab2032}.

\bibitem{Chaty+2022abn..book.....C}
S.~Chaty, \emph{Accreting Binaries}, 2514-3433 (IOP Publishing) (2022), \doi{10.1088/2514-3433/ac595f}, \url{https://dx.doi.org/10.1088/2514-3433/ac595f}.

\bibitem{Ivanova+2011ApJ...730...76I}
N.~{Ivanova}, {Common Envelope: On the Mass and the Fate of the Remnant}. \emph{\apj} \textbf{730}~(2), 76 (2011), \doi{10.1088/0004-637X/730/2/76}.

\bibitem{Wang+2009MNRAS.395..847W}
B.~{Wang}, X.~{Meng}, X.~{Chen}, Z.~{Han}, {The helium star donor channel for the progenitors of Type Ia supernovae}. \emph{\mnras} \textbf{395}~(2), 847--854 (2009), \doi{10.1111/j.1365-2966.2009.14545.x}.

\bibitem{Guo+2023MNRAS.526..932G}
Y.-L. {Guo}, \emph{et~al.}, {Type Ia supernovae in NS+He star systems and the isolated mildly recycled pulsars}. \emph{\mnras} \textbf{526}~(1), 932--941 (2023), \doi{10.1093/mnras/stad2578}.

\bibitem{Podsiadlowski+2001ASPC..229..239P}
P.~{Podsiadlowski}, {Common-Envelope Evolution and Stellar Mergers}, in \emph{Evolution of Binary and Multiple Star Systems}, P.~{Podsiadlowski}, S.~{Rappaport}, A.~R. {King}, F.~{D'Antona}, L.~{Burderi}, Eds., vol. 229 of \emph{Astronomical Society of the Pacific Conference Series} (2001), p. 239.

\bibitem{Passy+2012ApJ...744...52P}
J.-C. {Passy}, \emph{et~al.}, {Simulating the Common Envelope Phase of a Red Giant Using Smoothed-particle Hydrodynamics and Uniform-grid Codes}. \emph{\apj} \textbf{744}~(1), 52 (2012), \doi{10.1088/0004-637X/744/1/52}.

\bibitem{MacLeod+2015ApJ...798L..19M}
M.~{MacLeod}, E.~{Ramirez-Ruiz}, {On the Accretion-fed Growth of Neutron Stars during Common Envelope}. \emph{\apj} \textbf{798}~(1), L19 (2015), \doi{10.1088/2041-8205/798/1/L19}.

\bibitem{Manchester+2005AJ....129.1993M}
R.~N. {Manchester}, G.~B. {Hobbs}, A.~{Teoh}, M.~{Hobbs}, {The Australia Telescope National Facility Pulsar Catalogue}. \emph{\aj} \textbf{129}~(4), 1993--2006 (2005), \doi{10.1086/428488}.

\end{thebibliography}
\end{document}